\documentclass[aps,prc,showpacs,showkeys,superscriptaddress,nofootinbib,twocolumn,floatfix]{revtex4}
\usepackage{bm}
\usepackage{placeins}
\usepackage[]{graphicx}
\usepackage{amsfonts}
\usepackage{float}
\usepackage{multirow}
\usepackage{graphicx,color,amsmath,amssymb}

\usepackage[colorlinks=true, linkcolor=blue, citecolor=blue, urlcolor=blue]{hyperref}

\newcommand{\eg}{\textit{e.g.}}

\newcommand{\ie}{\textit{i.e.}}
\newcommand{\dd}{\mathrm{d}}
\newcommand{\lsim}{\lesssim}
\newcommand{\gsim}{\gtrsim}
\newcommand{\bc}{B_c}
\newcommand{\bcs}{B_c\left(1S\right)}
\newcommand{\bcp}{B_c\left(1P\right)}
\newcommand{\barb}{\bar{b}}
\newcommand{\raa}{R_{\rm AA}}
\newcommand{\npart}{N_{\rm part}}
\newcommand{\pT}{p_T}
\newcommand{\Td}{T_{\rm diss}}

\begin{document}

\title{Recombination of $B_c$ mesons in ultrarelativistic heavy-ion collisions}

\author{Biaogang Wu\footnote{bgwu@tamu.edu}}
\address{Cyclotron Institute and Department of Physics and Astronomy, Texas A$\&$M University, College Station, TX 77843-Interactions3366, USA}
\author{Zhanduo Tang\footnote{zhanduotang@tamu.edu}}
\address{Cyclotron Institute and Department of Physics and Astronomy, Texas A$\&$M University, College Station, TX 77843-3366, USA}
\author{Min He\footnote{minhephys@gmail.com}}
\address{Department of Applied Physics, Nanjing University of Science and Technology, Nanjing~210094, China}
\author{Ralf Rapp\footnote{rapp@comp.tamu.edu}}
\address{Cyclotron Institute and Department of Physics and Astronomy, Texas A$\&$M University, College Station, TX 77843-3366, USA}
\date{\today}

\begin{abstract}
High-energy heavy-ion collisions have been suggested as a favorable environment for the production of $\bc$ mesons, due to a much larger abundance of charm and bottom quarks compared to elementary reactions. Motivated by recent CMS data for $\bc^+$ production in Pb-Pb(5.02\,TeV) collisions at the
Large Hadron Collider (LHC), we deploy a previously developed transport approach for charmonia and bottomonia to evaluate the kinetics of $\bc$ mesons throughout the fireball formed in these reactions.
The main inputs to our approach are two transport parameters: the $\bc$'s reaction rate and equilibrium limit.
Both quantities are determined by previous calculations via a combination of charm and bottom sectors. In-medium binding energies of $\bc$ mesons are calculated from a thermodynamic $T$-matrix with a lattice-QCD constrained potential, and figure in their inelastic  reaction rates.
Temperature-dependent equilibrium limits include charm- and bottom-quark fugacities based on their initial production. 
We compute the centrality dependence of inclusive $\bc$ production and transverse-momentum ($\pT$) spectra using two different recombination models, instantaneous coalescence and resonance recombination. The main uncertainty
in the resulting nuclear modification factors, $\raa$, is currently associated with the $\bc$ cross section in elementary $pp$ collisions, caused by the uncertainty in the branching ratio for the
$\bc^-\to J/\psi\mu^-\bar \nu$ decay. 
Our results indicate a large enhancement of the $\raa$ at low $\pT$, with significant regeneration contributions up to $\pT\simeq$\,20\,GeV. 
Comparisons to CMS data are carried out but firm conclusions will require a more accurate value of the branching ratio, or alternative channels to measure the $\bc$ production in $pp$ collisions.

\end{abstract}
\keywords{Quark-Gluon Plasma, Heavy Quarkonia, Ultra-relativistic Heavy-Ion Collisions}
\maketitle

\section{Introduction}
\label{sec_intro} 
Measurements of charmonia in ultra-relativistic heavy-ion collisions (URHICs) at the Large Hadron Collider (LHC) have demonstrated the importance of charm-quark recombination processes 
in the strongly interacting fireball formed in these reactions~\cite{Andronic:2015wma}. While the microscopic description and precise magnitude of recombination contributions remain under some debate~\cite{Andronic:2014sga,Zhou:2016vwq,Scomparin:2017pno,Rapp:2017chc}, the measured dependence of $J/\psi$ production on 
collision centrality (with an approximately constant nuclear modification factor, $\raa$), transverse momentum (being concentrated at low $\pT$) and azimuthal emission angle (with a sizable elliptic flow, $v_2$), give strong evidence for recombination of nearly thermalized charm ($c$) and anticharm ($\bar c$) quarks in the fireball. In the bottomonium sector, this 
evidence is less pronounced, although transport calculations predict a non-negligible component of regeneration in an overall suppressed $\raa$ of $\Upsilon$ mesons in Pb-Pb collisions~\cite{Du:2017qkv,Yao:2018sgn}.
The ratio of bottomonium over total bottom production in $pp$ collisions of typically a few permille is much smaller than the $\approx$ 1\% for charmonia.
It is therefore of great interest to study bound states of bottom ($b$) and $c$ quarks, \ie, $\bc^+$ mesons.
For the ground state, $\bc(6275)$, the production fraction in $pp$ collisions relative to $b \barb$ has recently been reported at $\approx$ 0.25\,\%~\cite{LHCb:2019tea}, with a significant uncertainty from theoretical calculations of the branching ratio for the $B_c^-\to J/\psi \mu^-\bar \nu$ decay. 
This suggests that $\bc$ formation via recombination of a $b$ ($\barb$) quark with the rather abundant $\bar c$ ($c$) quarks in Pb-Pb collisions at the LHC can be quite sizable relative to the $pp$ reference.

In a broader context, $\bc$ production is part of the program of using heavy quarkonia as a probe of the quark-gluon plasma (QGP) in URHICs~\cite{Rapp:2008tf,BraunMunzinger:2009ih,Kluberg:2009wc,Mocsy:2013syh,Zhao:2020jqu}, specifically to 
understand how their binding and kinetics are affected by the in-medium potential of quantum chromodynamics (QCD). The $\bc$ states open a new perspective on that, and also establish relations between the in-medium spectroscopy of charmonia and bottomonia.    
Originally discovered in $p\bar p$ collisions at Fermilab~\cite{CDF:1998axz}, $\bc$ mesons are now becoming accessible in URHICs. 
Pioneering data by the CMS Collaboration~\cite{CMS:2022sxl} indeed give a hint that $\bc^+$ production in Pb-Pb collisions is enhanced relative to expectations from  $pp$ collisions, currently measured with a restriction on $\pT>6$\,GeV. 
Earlier theoretical studies of $\bc$ production~\cite{Schroedter:2000ek,Liu:2012tn} predicted a large increase in their abundance relative to $pp$ collisions. For example, in Ref.~\cite{Liu:2012tn}, the $\bc$ nuclear modification factor was found to reach values of $\approx$ 2.5-17 at low $\pT$, depending on the assumption of the underlying in-medium heavy-quark (HQ) potential (free vs.~internal energy of the HQ pair), while the three-momentum dependence was assumed to be given by thermalized $\bc$ spectra.
More recently, an instantaneous coalescence model (ICM)~\cite{Chen:2021uar} was employed to 
calculate the yield of $\bc$'s
at a fixed temperature using bottom- and charm-quark distributions from Langevin transport simulations. The production yields in ICMs can be rather sensitive to the model for the underlying Wigner distribution functions, in particular to the spatial radius, which in Ref.~\cite{Chen:2021uar} was estimated using the free-energy potential.

In the present paper, we employ a kinetic rate equation~\cite{Grandchamp:2003uw,Zhao:2011cv,Du:2017qkv,Wu:2020zbx} to compute the time evolution of $\bc$ ($\bc^+$ and $\bc^-$) production for QGP fireballs in Pb-Pb collisions at the LHC.
The in-medium binding energies are determined from thermodynamic $T$-matrix calculations of $\bc$ spectral functions employing the strongly coupled QGP scenario of Ref.~\cite{Liu:2017qah}, with a potential extracted from thermal lattice-QCD (lQCD) 
data, which is much stronger than the HQ free energy. The latter has been shown to be incompatible with bottomonium data at the LHC~\cite{Emerick:2011xu,Strickland:2011aa}.  
With the resulting reaction rates and equilibrium limits, we calculate the centrality dependence of inclusive $\bc$ production including feeddown contributions from excited states.  
In the context of the CMS data, good control over the $\pT$ dependence of the yields is required, especially for the recombination contribution (which turns out to be large also in our calculation). 
Since the aforementioned $\pT>6$\,GeV cut employed by CMS is close to the $\bc$ mass, one is rather sensitive to the concrete implementation of the recombination processes whose $\pT$-dependence can vary considerably, \eg, through the inputs for the $c$- and $b$-quark spectra~\cite{Zhao:2022auq}.
We will therefore investigate the results for both an ICM and the resonance recombination model (RRM)~\cite{Ravagli:2007xx,He:2011qa}, thereby using state-of-the-art transported HQ spectra~\cite{He:2019vgs} that give a fair description of open heavy-flavor (HF) observables in Pb-Pb collisions at the LHC~\cite{ALICE:2021rxa}.  

This paper is organized as follows. In Sec.~\ref{sec_spec} we compute in-medium spectral functions 
of $S$- and $P$-wave $\bc$ states within the thermodynamic $T$-matrix approach, extract their binding energies and calculate pertinent reaction rates in the QGP. 
In Sec.~\ref{sec_kin} we introduce the kinetic-rate equation and evaluate 
its second transport parameter, the $\bc$ equilibrium limit, including its 
dependence on the cross section inputs for open HF production in $pp$ collisions and their shadowing corrections. 
In Sec.~\ref{sec_prod} we study the time dependence of the $\bc$ kinetics and discuss the resulting centrality dependence of the $\raa$ for inclusive $\bc$ production in Pb-Pb(5.02\,TeV) collisions. 
In Sec.~\ref{sec_pt} we detail the calculations of the $\bc$'s $\pT$ spectra using two different recombination models. This allows us to extract the centrality dependent $\raa$ with a $\pT>6$\,GeV cut and compare it to CMS data.
In Sec.~\ref{sec_concl} we summarize our work and conclude.

\section{$\bc$ Spectral Functions in the QGP}
\label{sec_spec} 
When utilizing quarkonia as a probe of the QGP, their in-medium spectral properties play a key role in determining transport parameters that are required to compute observables suitable for comparison with the experiment. While this program has been widely carried out for charmonia and bottomonia, we are not aware of microscopic calculations of in-medium spectral functions of $\bc$ mesons to date. Toward this end, we employ a thermodynamic $T$-matrix approach along the lines of previous investigations~\cite{Mannarelli:2005pz,Cabrera:2006wh,Riek:2010fk,Liu:2017qah}. 
It is based on a temperature-dependent two-body potential and solved 
self-consistently for the resummed Dyson-Schwinger equations of the in-medium one- and two-parton correlation functions in the QGP, schematically written as
\begin{eqnarray}
T_{Q\bar Q} &=& V_{Q\bar Q} + \int \dd k V_{Q\bar Q} D_{Q}(k) D_{\bar Q}(p-k) T_{Q\bar Q}   
\\
D_{Q}(k) &=&  1/[k_0-\omega_{1,k} -\Sigma_1(k)]  
\\
\Sigma_Q(k) &= & \int \dd p T_{Qi} D_{i}(p) f_i\ ,
\end{eqnarray}
where $T_{Q\bar Q}$ denotes the quarkonium $T$-matrix, $D_{Q,i}$ single-parton propagators for either heavy quarks ($Q$) or thermal partons ($i=q,\bar q,g$), and $f_i$ the pertinent thermal-parton distribution function. 
The input potential is taken of Cornell type with in-medium screened color-Coulomb and string interactions, which in the color-singlet amounts to the ansatz
\begin{equation}
   V(r;T) = -\frac{4}{3} \alpha_{s} \left[\frac{e^{-m_{d} r}}{r} + m_{d}\right] -\frac{\sigma}{m_s} \left[e^{-m_{s} r-\left(c_{b} m_{s} r\right)^{2}}-1\right]
\ .
\end{equation}
A Fourier transform into momentum space is carried out, followed by a partial-wave expansion of the $T$-matrix equation. We adopt the parameters of Ref.~\cite{Liu:2017qah}, where the coupling constant and string tension are fixed at $\alpha_s=0.27$ and $\sigma=0.225$ $\textup{GeV}^2$, respectively, to 
reproduce lQCD data for the HQ free energy in vacuum. 
The finite-temperature screening masses, $m_{d,s}$, are related via  $m_s=\left(c_{s} m_{d}^2 \sigma/\alpha_s\right)^{1/4}$, where $m_d(T)$ and $c_s$ are parameters of the in-medium potential, while $c_b$ controls the string-breaking distance. Together with the bare masses for the light thermal partons, they are used to fit the selfconsistent $T$-matrix results to finite-temperature lQCD data for the HQ free energy, Euclidean quarkonium correlator 
ratios, and the equation of state of the QGP. Here, we focus on a solution referred to as a strongly coupled scenario (SCS)~\cite{Liu:2017qah}. 
Compared to the solution of a weakly coupled scenario (WCS), the SCS is preferred by yielding transport parameters~\cite{Liu:2016ysz} that are close to the
ones extracted from phenomenological studies based on hydrodynamics and HQ transport models~\cite{He:2022ywp}.  
\begin{figure}[t]
   \begin{minipage}[b]{1\linewidth}
        \centering
        \includegraphics[width=0.9\textwidth]{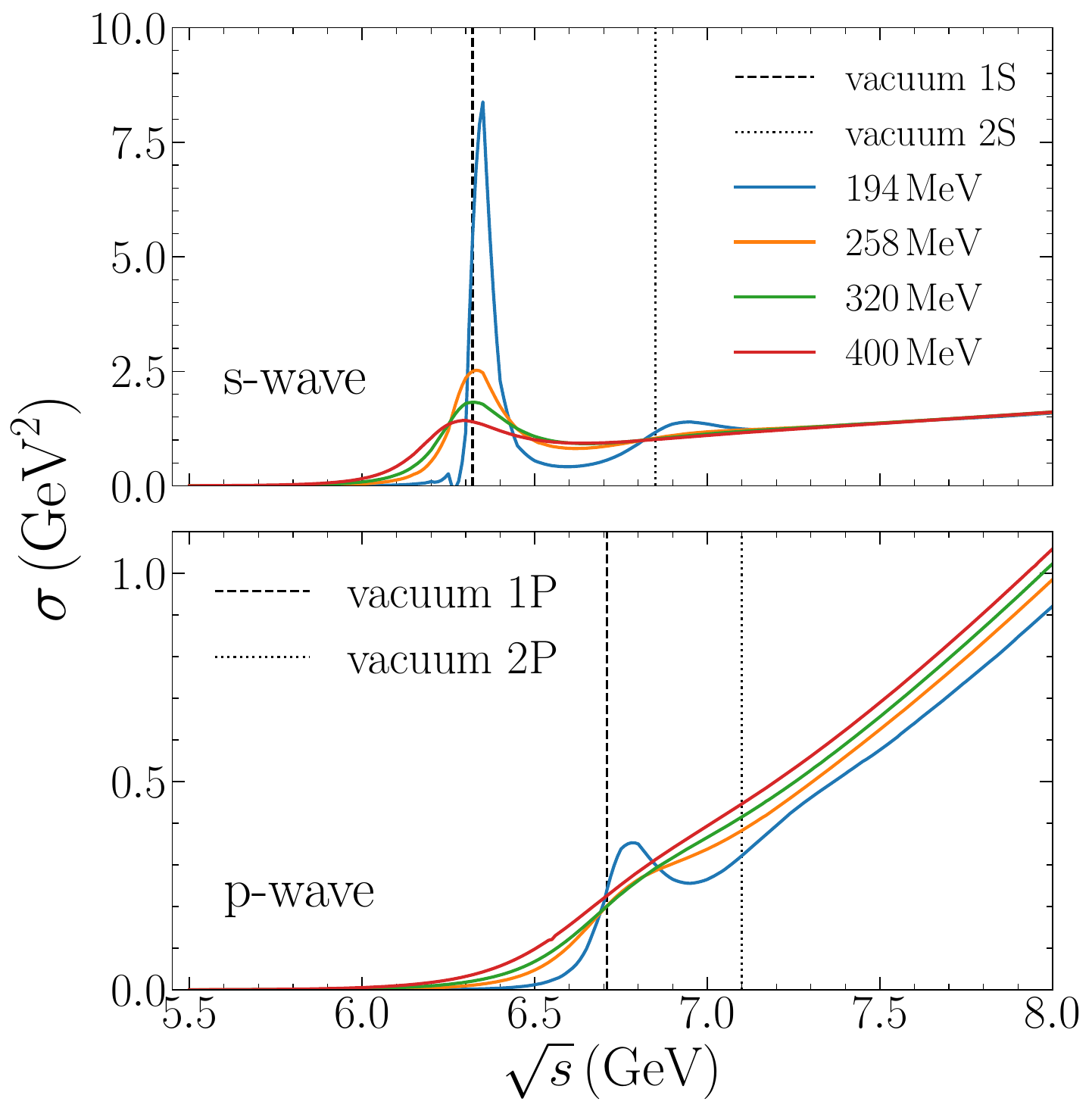}
 \end{minipage}
 \caption{In-medium spectral functions of $S$- (upper panel) and $P$-wave (lower panel) $\bc$ channels at different temperatures. The vacuum masses of $\bcs$ and $\bcp$ states are indicated by the thick- (ground states) and thin-dashed (excited states) vertical lines.
 }
\label{fig_sf}
\end{figure}

Focusing now on the heavy-quarkonium sector, we first note that the vacuum charmonium and bottomonium ground-state masses
can be reproduced with a string-breaking distance of 
$r_{SB}=1.1$\,fm in connection with constituent HQ masses given by 
$m_Q={V}(\infty)/2+m_Q^0$ (where ${V}(\infty)$ denotes the potential value at an infinite distance) and bare masses of
$m_{c,b}^0=1.264, 4.662$\,GeV~\cite{Riek:2010fk}.
With this setup, the results for the $\bc$ spectral functions follow without further parameters or assumptions. The vacuum spectrum for both $S$ and $P$ 
states is shown in Fig.~\ref{fig_sf} by the dashed vertical lines. Since we do not account for fine or hyperfine splittings (which are of higher order in $1/m_Q$), the ($S$-wave) pseudoscalar and vector channels are degenerate. 
The calculated vacuum masses of $\bc$ (6.324~GeV) and $\bc(2S)$ (6.850~GeV) are in approximate agreement with the experimentally measured values of 6.274\,GeV for the pseudoscalar ground state and 6.871\,GeV for its putative 2$S$ excitation, respectively, which are the only known ones thus far~\cite{ParticleDataGroup:2022pth}. We also predict two $P$-wave $\bc$ bound states,  $\bcp$ and $\bc(2P)$, with masses 6.711 and 7.100~GeV, respectively.
\begin{figure}[t]
\begin{minipage}[b]{01.00\linewidth}
        \centering
        \includegraphics[width=0.9\textwidth]{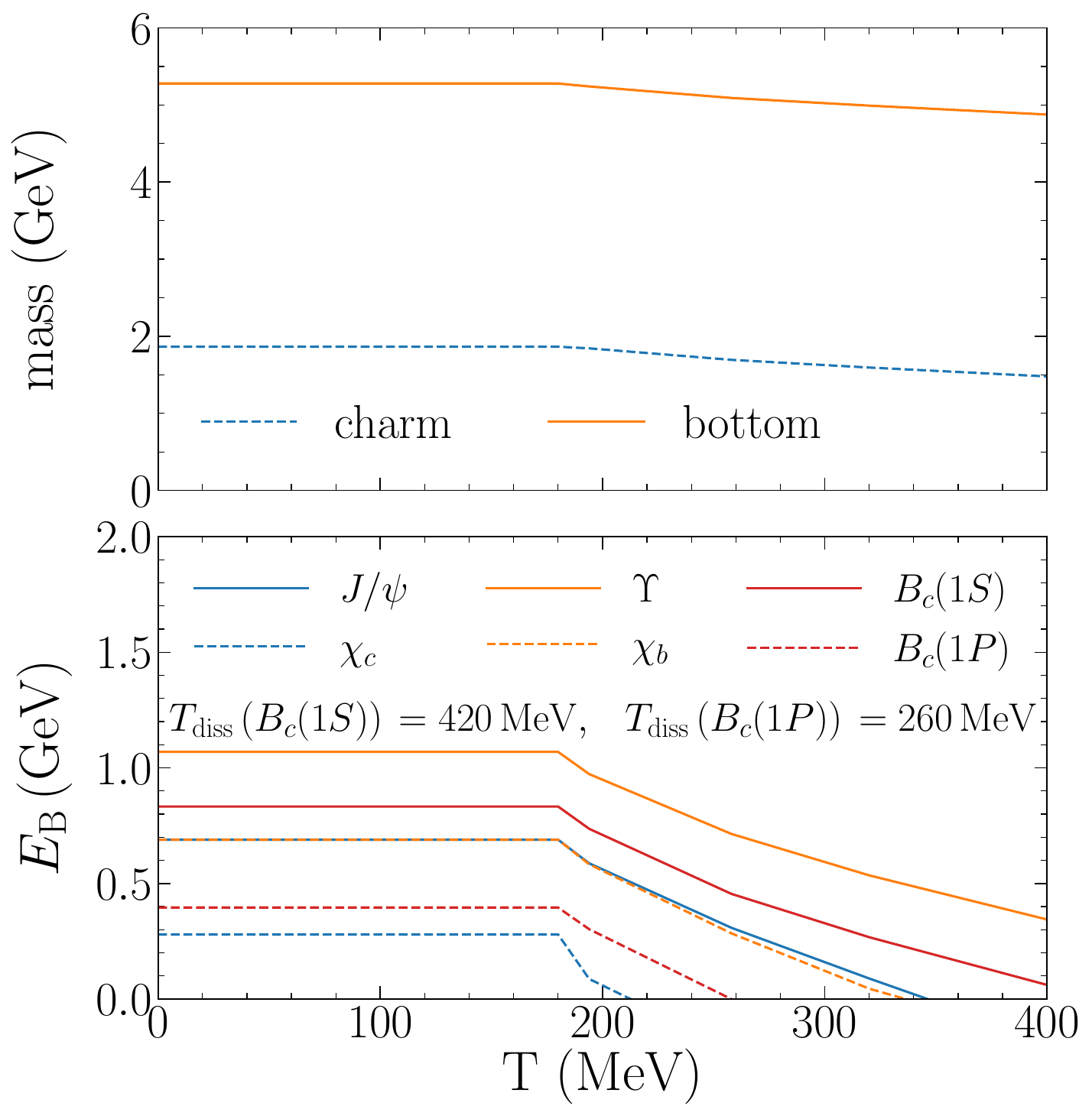}
     \end{minipage}
     \caption{Charm- and bottom-quark masses (upper panel) and binding energies of $\bcs$, $\bcp$, $J/\psi$, and  $\Upsilon(1S)$ (lower panel) as a function of temperature, as obtained from the $T$-matrix approach.}
       \label{fig_BE}
\end{figure}

\begin{figure}[t]
\begin{minipage}[b]{01.00\linewidth}
   \includegraphics[width=0.98\textwidth]{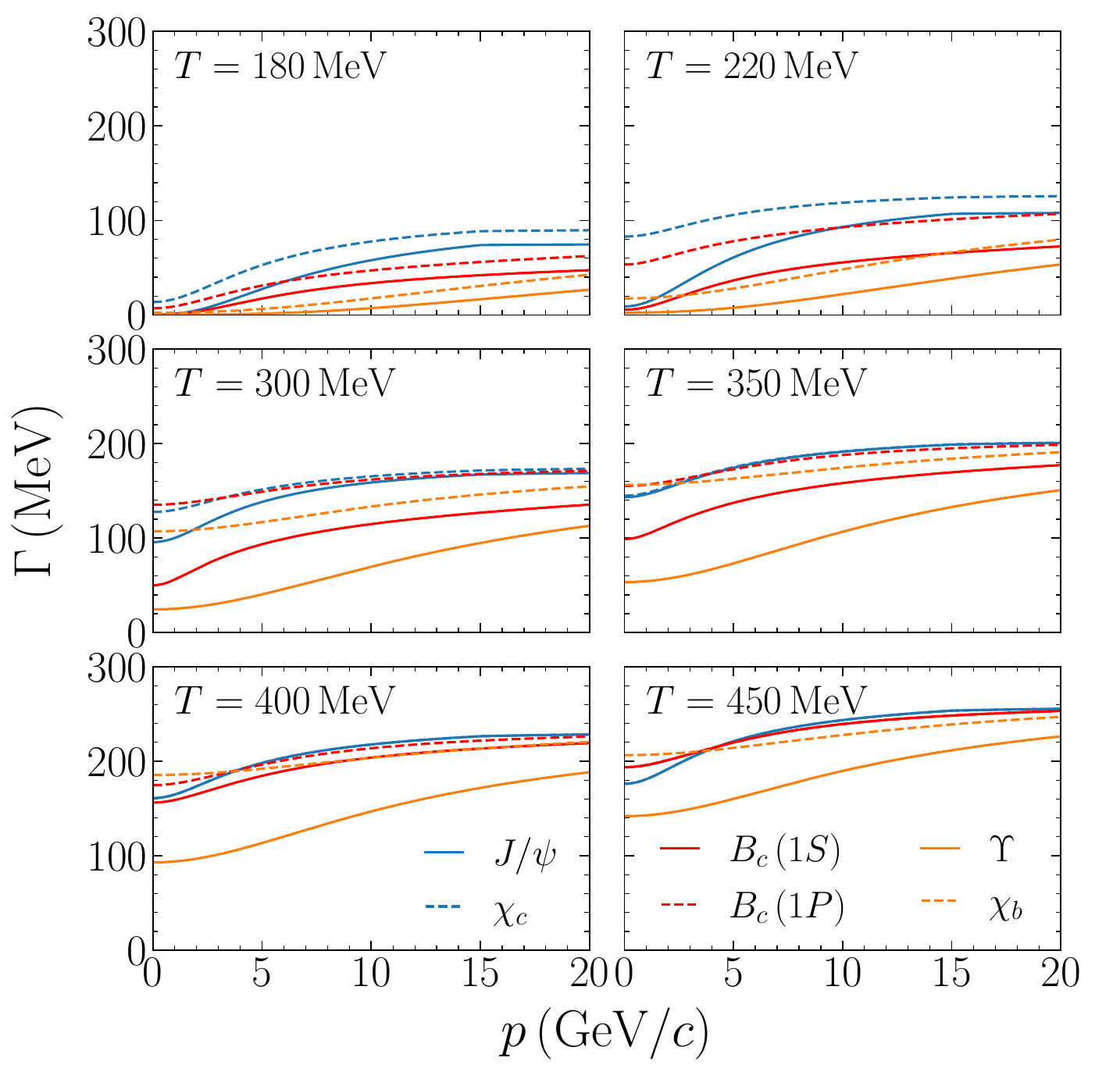}
     \end{minipage}
     \caption{Quasifree reaction rates of charmonia, bottomonia and $\bc$ states as a function of momentum at different temperatures. }
\label{fig_rate}
\end{figure}

The $S$- and $P$-wave spectral functions in the QGP are calculated by closing the two incoming and outgoing legs of the $T$-matrix, plus a non-interacting continuum independent of the $T$-matrix, with the corresponding projection operators for the different channels. They are also shown in Fig.~\ref{fig_sf}.
The spectral functions broaden with increasing temperature, indicating the gradual dissociation of the bound states. The $S$-wave ground state survives to rather high
temperatures of $T\simeq400$ MeV, while the $P$-wave ground state ceases to exist for temperatures of 
$T\gsim 250$\,MeV. The dissolution of the $\bc$ states results from the large scattering rates of charm and 
bottom quarks in the medium, together with an increase in the screening of in-medium potentials at higher temperatures. 
The in-medium $\bcs$ mass turns out to be rather constant with temperature, due to a nontrivial interplay of decreasing HQ masses and binding energy, similar to what has been found for charmonia~\cite{Liu:2017qah}.
\begin{figure}[t]
\begin{minipage}[b]{01.00\linewidth}
        \centering
        \includegraphics[width=0.9\textwidth]{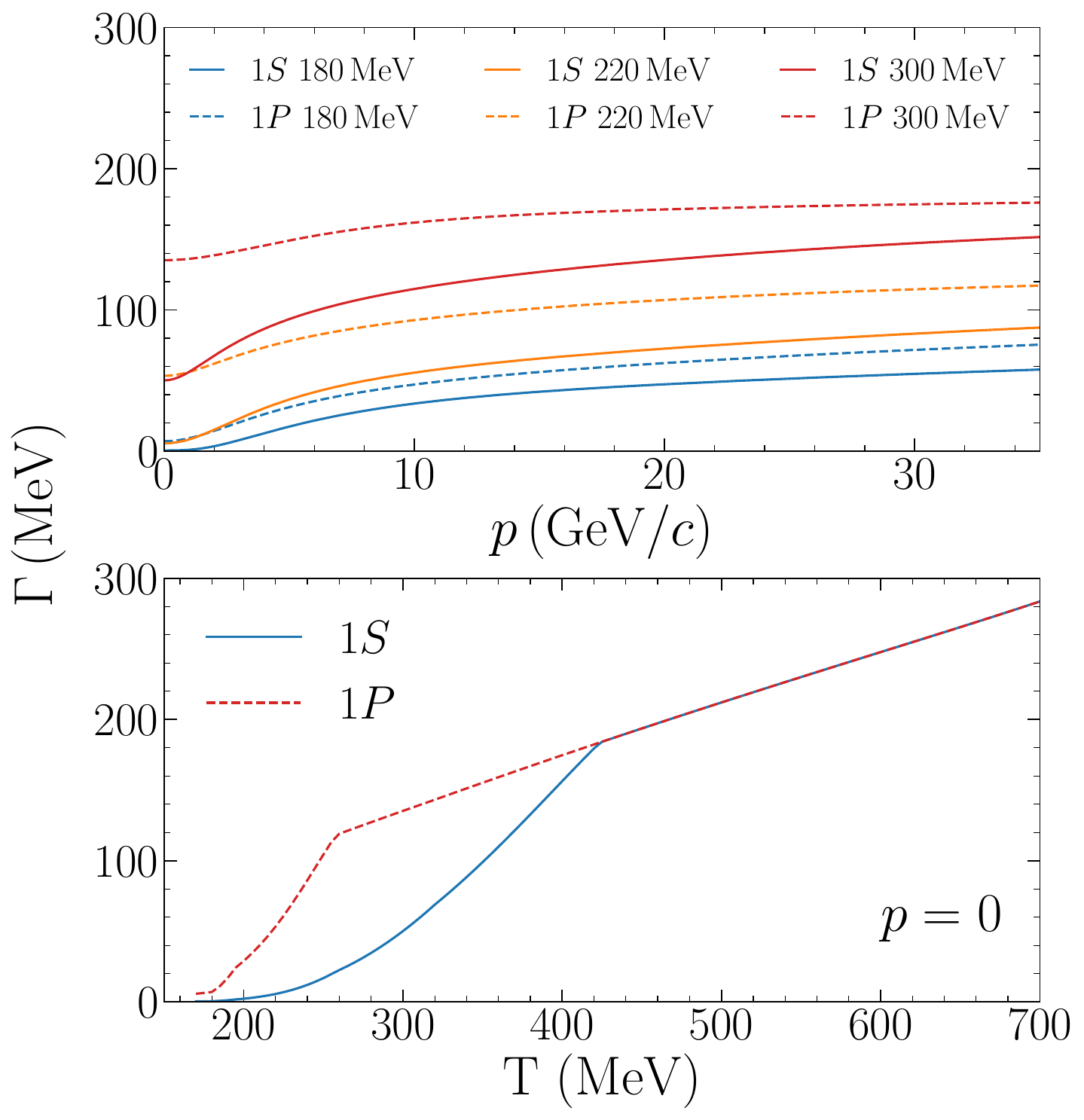}
     \end{minipage}
        \caption{Parton-induced quasifree dissociation rates for $\bcs$ (solid lines) and $\bcp$ (dashed lines)  
        at $T$=180\,MeV, $T$=220\,MeV, and $T$=300\,MeV (upper panel) and as a function of temperature for vanishing three-momentum (lower panel). We note that here and in Fig.~\ref{fig_rate} the rates are plotted beyond the temperature where the corresponding binding energy vanishes where they simply become the sum of the constituent HQ scattering rates (simulating the destruction of a would-be bound state correlation).}
   \label{fig_rate_Bc}
\end{figure}
From the spectral functions, we can extract the in-medium binding energies of different $\bc$ states which we define as the difference between the
nominal in-medium charm- plus bottom-quark masses and the peak position of a given state. The former are shown in the upper 
panel of Fig.~\ref{fig_BE} and the (magnitude of the) binding energies, $E_B$, in the lower panel. The latter essentially retain the vacuum hierarchy of charmonia, bottomonia and $\bc$ binding. Following earlier work 
within the TAMU quarkonium transport model~\cite{Zhao:2010nk}, we employ the in-medium binding energies from the SCS to 
calculate the inelastic reaction rates of the $\bc$ states. The dominant contribution arises from inelastic scatterings 
of thermal partons ($i$=$q,\bar{q},g$) off the heavy quarks inside the bound state, $i+\bc \to c+\bar{b} +i$ 
(this even holds for the more strongly bound bottomonia~\cite{Du:2017qkv}).  
We implement these processes in the so-called quasifree approximation, where the inelastic reaction is calculated through 
half-off-shell scattering on either heavy quark in the bound state whose virtuality accounts for the binding energy while the other quark is treated as a spectator (this amounts to neglecting recoil corrections while four-momentum is 
conserved)~\cite{Grandchamp:2001pf}. The dissociation rate for $\bc$ takes the form
\begin{eqnarray}
   \Gamma_{\bc}^{\rm qf}(p,T)&=&\sum\limits_i\int\frac{\dd^3p_{i}}{(2\pi)^3} f_{i}(\omega_{p_i},T) 
\nonumber\\
&& \times [v^{ic} \sigma_{ic\rightarrow ic}(s)  + v^{i\barb} \sigma_{i\barb\rightarrow i\barb}(s) ]\ ,
\label{eq-quasifree}
\end{eqnarray}
where $f_i$ are thermal parton distribution functions 
(Fermi or Bose), $s=(p_Q+p_i)^2$, and
\begin{equation}
v_{Qi}=\frac{\sqrt{\left(p_{Q}^{(4)} \cdot p_{i}^{(4)}\right)^{2}-m_{Q}^{2} m_{i}^{2}}}{\omega_{Q}(p_Q) \omega_{i}(p_i)}
\end{equation}
is the relative velocity of the incoming $b$ or $c$ quark and a thermal parton. Figure~\ref{fig_rate_Bc}
shows the quasifree dissociation rates for $\bcs$ and $\bcp$ as a function of their three-momentum, $p$, for various 
temperatures (upper panel) and as a function of the temperature at $p$=0.
Their main features are an 
increase with three-momentum that is more pronounced for large binding, a significant decrease with 
increasing binding energy (comparing ground and excited states), and a marked overall increase with temperature.
In Fig.~\ref{fig_rate} we show a systematic comparison of the reaction rates for the various $S$- and $P$-wave calculated with the in-medium binding energies shown in Fig.~\ref{fig_BE}, using the same framework as in previous works for charmonia~\cite{Zhao:2010nk} and bottomonia~\cite{Du:2017qkv} (for simplicity we do not include interference effects). 
As expected, the dissociation rates for $\bc$ states generally 
lie in between the ones of the corresponding charmonium and bottomonium states,
with the exception of near-vanishing binding and low momentum, which is presumably caused by different 
recoil kinematics for $b$ and $c$ quarks,

\section{Kinetic Approach}
\label{sec_kin} 
The kinetic rate equation for the number, $N_{\bc}$, of a specific $\bc$ state is given by~\cite{Grandchamp:2003uw,Zhao:2010nk}
\begin{equation}
   \frac{\dd N_{\bc}(\tau)}{\dd\tau}=-\Gamma_{\bc}(T(\tau))\left[N_{\bc}(\tau)-N_{\bc}^{\rm eq}(T(\tau))\right] \ .
\label{rate-eq}
\end{equation}
In the present work, we will consider $\bcs$ and $\bcp$ states.
The rate equation requires two transport parameters; the equilibrium limit $N_{\bc}^{\rm eq}$, 
and the reaction rate $\Gamma_{\bc}$.
%
The equilibrium limit is calculated from the statistical model, taking the form
\begin{equation}
   N_{\bc}^{\rm eq}(T) =  V_{\rm FB} d_{\bc} \gamma_c\gamma_{\barb} \int \frac{\dd^3k}{(2\pi)^3} \exp(-E_k/T)     \  , 
\end{equation}
where $E_k=\sqrt{k^2+m_{\bc}^2}$, and $V_{\rm FB}$ is the time-dependent volume of the expanding fireball. 
We neglect the spin-induced $1/m_Q$ corrections in this work. Therefore, using standard spectroscopic notation, $^{2S+1}{L}_{J}$, the four $S$-wave states 
$^{1}{S}_{0}$ and $^{3}{S}_{1}$ are degenerate, and so are the 12 $P$-wave states $^{3}{P}_{0}$, $^{1}{P}_{1}$, $^{3}{P}_{1}$ and $^{3}{P}_{2}$
(the individual degeneracy of each state is given by a factor of $2J+1$). 
The equilibrium limits critically depend on the fugacity factors $\gamma_c$ and $\gamma_{\barb}$, which have been computed in our earlier works~\cite{Zhao:2010nk,Du:2017qkv}  assuming  conservation of $b\barb$ and $c\bar{c}$ pairs throughout the fireball expansion,
\begin{equation}
\label{Neq}
N_{Q\bar Q}=\frac{1}{2}\gamma_{Q} n_{\rm{op}}V_{\rm{FB}}\frac{I_1(\gamma_{Q} n_{\rm{op}}V_{\rm{FB}})}
{I_0(\gamma_{Q} n_{\rm {op}}V_{\rm{FB}})} + \gamma_{Q}^2 n_{\rm{hid}} V_{\rm{FB}}  \ , 
\end{equation}
where $I_0$ and $I_1$ are the modified Bessel functions of the zeroth and first order. The open ($n_{\rm{op}}$) and hidden ($n_{\rm{hid}}$) charm densities are matched to the number of charm-anticharm and bottom-antibottom quark pairs, $N_{c\bar{c}}$ and $N_{b\barb}$, produced
in primordial nucleon-nucleon collisions of the heavy-ion system (including shadowing corrections detailed below). 
\begin{figure}[t]
\begin{minipage}[b]{01.00\linewidth}
   \includegraphics[width=0.9\textwidth]{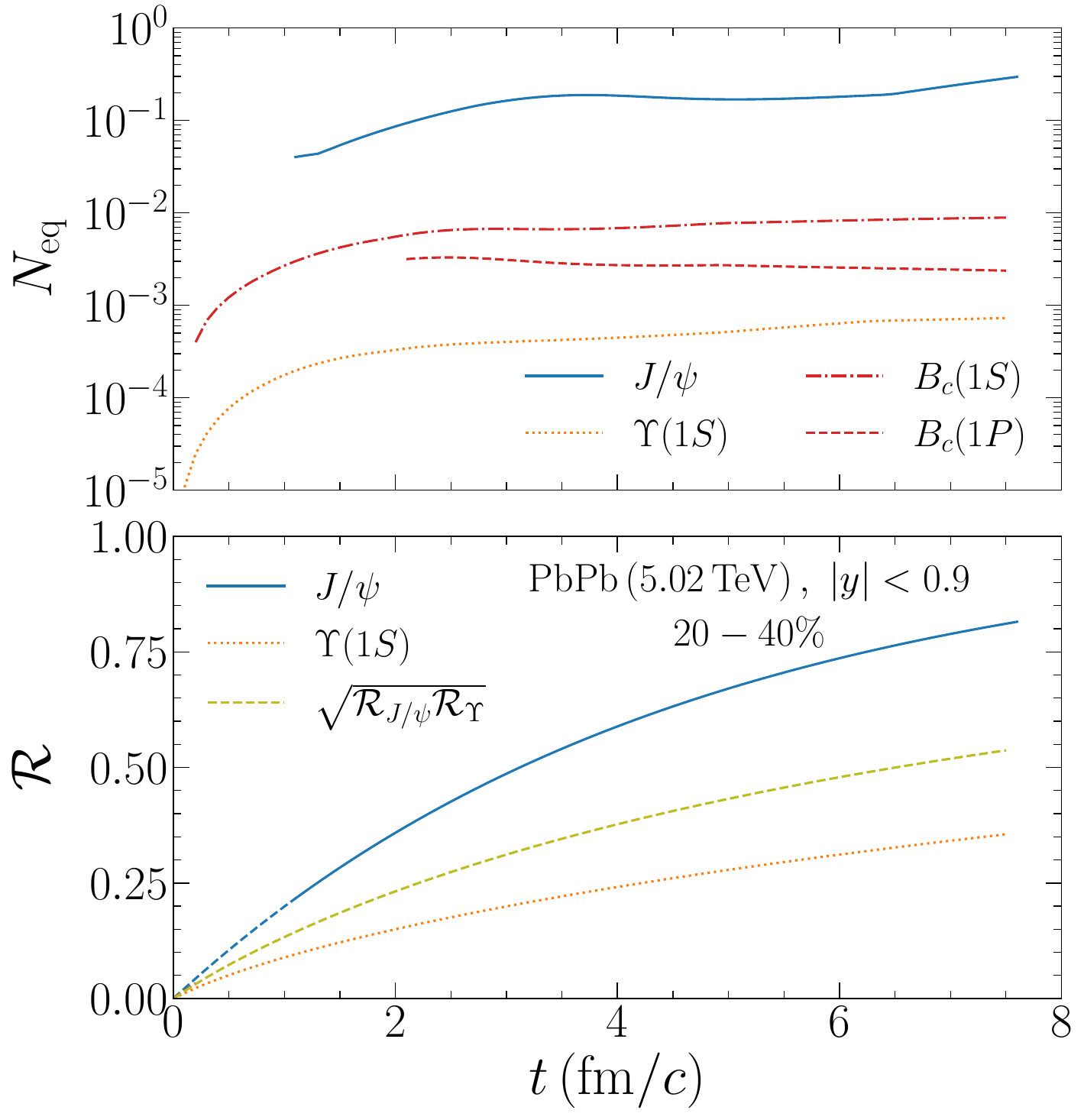}
     \end{minipage}
     \caption{Equilibrium limits of $\bcs$, $\bcp$, $J/\psi$ and  $\Upsilon(1S)$ (upper panel) and thermal-relaxation time factors (lower panel) as 
    a function of time in 20-40\% central 5.02\,TeV Pb-Pb collisions.
     }
\label{fig_EQ}
\end{figure}
In the QGP phase, the degrees of freedom are charm and bottom quarks for the open HF states (quarkonium contributions are negligible).
The fugacities of $b$ and $c$ quarks are quite different at low temperatures, due to the large difference in 
mass (compared to temperature); \eg, at $T$=200\,MeV, one has $\gamma_c$= 13.8 and $\gamma_b= 3.7 \times 10^6$. 
The time evolution of the equilibrium limits in minimum-bias Pb-Pb(5.02\,TeV) collisions of $\bcs$, $\bcp$ are displayed the in upper panel of Fig.~\ref{fig_EQ} and compared to those of $J/\psi$ and $\Upsilon(1S)$. 
%

The equilibrium limits given by Eq.~(\ref{rate-eq}) are valid when the heavy quarks are thermalized.
In the early stages of the fireball, the heavy quarks produced in the collision cannot be expected to be kinetically equilibrated; this generically leads to a suppression of quarkonium equilibrium limits as harder HQ distributions are less favorable for quarkonium formation than in the thermalized case~\cite{Grandchamp:2002wp,Song:2012at,Du:2022uvj}. 
We account for this effect as before~\cite{Grandchamp:2002wp,Grandchamp:2005yw} in a relaxation time approximation, by combining the effects of $c$ and $b$ quarks.
The pertinent relaxation time factor for a heavy quarkonium, $\mathcal{Q}=Q\bar{Q}$, is defined by~\cite{Grandchamp:2002wp}
\begin{equation}
   \mathcal{R}_{\mathcal{Q}}(t)=1-\exp \left(-\int_0^t \frac{\mathrm{d} t^{\prime}}{\tau_{\mathcal{Q}}\!\left(T\left(t^{\prime}\right)\right)}
   \right) \ ,
\end{equation}
with previously employed values of a constant $c$-quark relaxation time $\tau_c\simeq4.5\,{\rm fm}/c$~\cite{Zhao:2010nk} and a $b$-quark relaxation time decreasing with increasing temperature, $\tau_b\simeq11\,{\rm fm}/c$ at $\approx 2T_c$~\cite{Du:2017qkv} for charmonia and bottomonia, respectively. 
To infer the relaxation time factor for $\bc$, 
we first note that the equilibrium number of the $\bc$ can be approximately written as $N_{B_c}^{\mathrm{eq}}\sim\sqrt{N_{b\bar{b}}^{\mathrm{eq}} N_{c\bar{c}}^{\mathrm{eq}}}$. This follows from the approximate relation $m_{B_c} \simeq (m_{J/\psi}+m_{Y(1S)})/2$ and is also in line with relative chemical equilibrium as $\gamma_{B_c} = \gamma_c\gamma_b$. 
Consequently, we employ the following thermal relaxation factor for $\bc$:
\begin{equation}
\mathcal{R}_{b\bar{c}}(t)=\sqrt{\mathcal{R}_{b\bar{b}}(t)\mathcal{R}_{c\bar{c}}(t)} \ , 
\end{equation}
which is mostly governed by the slower relaxation of $b$ quarks.
%
A comparison of the thermal relaxation factors is depicted in the lower panel of Fig.~\ref{fig_EQ}, and they are also included in the equilibrium limits plotted in the upper panel.
%
\begin{figure}[t]
\begin{minipage}[b]{1.0\linewidth}
\centering
\includegraphics[width=0.9\textwidth]{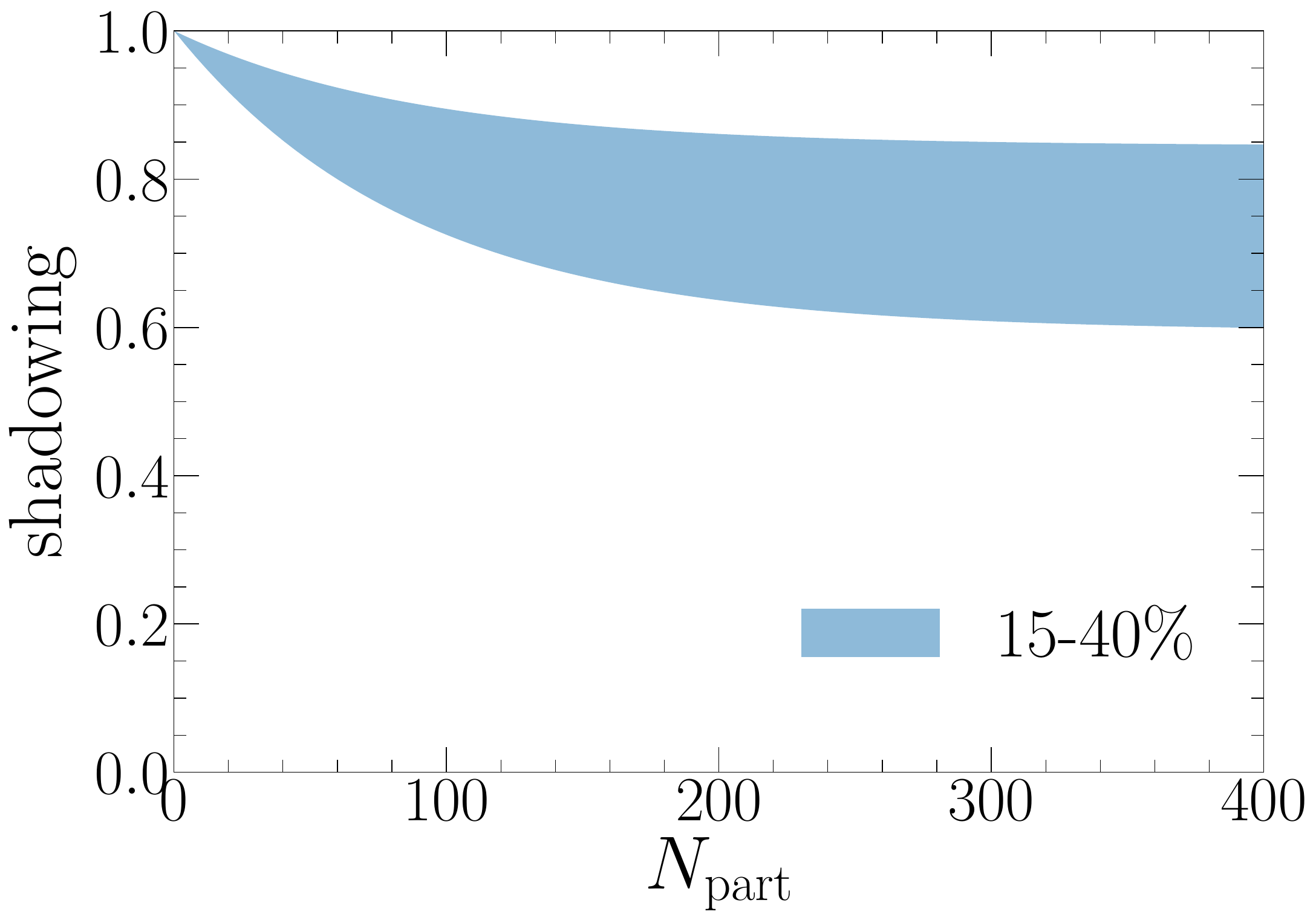}
\end{minipage}
\caption{Centrality dependence of shadowing for $\bc$ in 5.02\,TeV Pb-Pb collisions. The uncertainty of the total shadowing in most central collisions is 10-30\%.
} 
\label{fig_shadow}
\end{figure}

The numbers of HQ pairs, $N_{Q\bar{Q}}$, are calculated from their production cross sections in proton-proton 
($pp$) collisions times the number of primordial nucleon-nucleon collisions $N_{\rm Coll}$, as estimated from 
the optical Glauber model for heavy-ion collisions at given centrality (and energy). In 5.02\,TeV $pp$ collisions, we use recent ALICE measurements at mid-rapidity, \ie,
$\dd\sigma_{c\bar c}/\dd y=1.15$\,mb~\cite{ALICE:2021dhb},  
and $\dd \sigma_{b\barb}/\dd y|_{\left|y\right|<0.5}=34.5\pm2.4^{+4.7}_{-2.9}\,\mu$b~\cite{ALICE:2021mgk}. 
An $\npart$ dependent shadowing is applied which suppresses the total $c\bar c$ ($b\barb$) cross-section by up to 10(0)-30(10)\% in most central collisions. 

To compute observables, usually presented in terms of a nuclear modification factor [see Eq.~(\ref{RAA-eq}) below],
we also need the production cross section of $\bc$ states in $pp$ collisions. Measurements are currently restricted to the 
$\bcs$ in the $J/\psi\mu\nu$ decay channel, quoted as~\cite{LHCb:2019tea}.
\begin{equation}
    \begin{aligned}
    \frac{\sigma_{\bc^-}}{\sigma_{b\barb}} \mathcal{B}( \bc^- \rightarrow &J/\psi\mu^- \bar{\nu})\\
    =&\left(5.04\pm0.11\pm0.17\pm0.18\right)\times 10^{-5}
\end{aligned}
\end{equation}
The pertinent branching ratio, 
$\mathcal{B}\left( \bc^- \rightarrow J/\psi\mu^- \bar{\nu} \right)$, has been evaluated in various theoretical 
models providing a range of ${\cal B}\simeq 1.3-7.5$\%; cf.~the compilation in
Ref.~\cite{LHCb:2019tea}. Here, we take the median value of these calculations (excluding the lowest and highest value from the list) as our best estimate with a $1\sigma$ (68\%) confidence range of the number of models around the median, 
to obtain ${\cal B}=1.9^{+0.7}_{-0.4}\%$; this translates into $\dd \sigma_{\bc\left(1S\right)}/\dd y=57.8-110.8$\,nb.
%
\begin{figure}[t]
\begin{minipage}[b]{1.0\linewidth}
\centering
\includegraphics[width=0.9\textwidth]{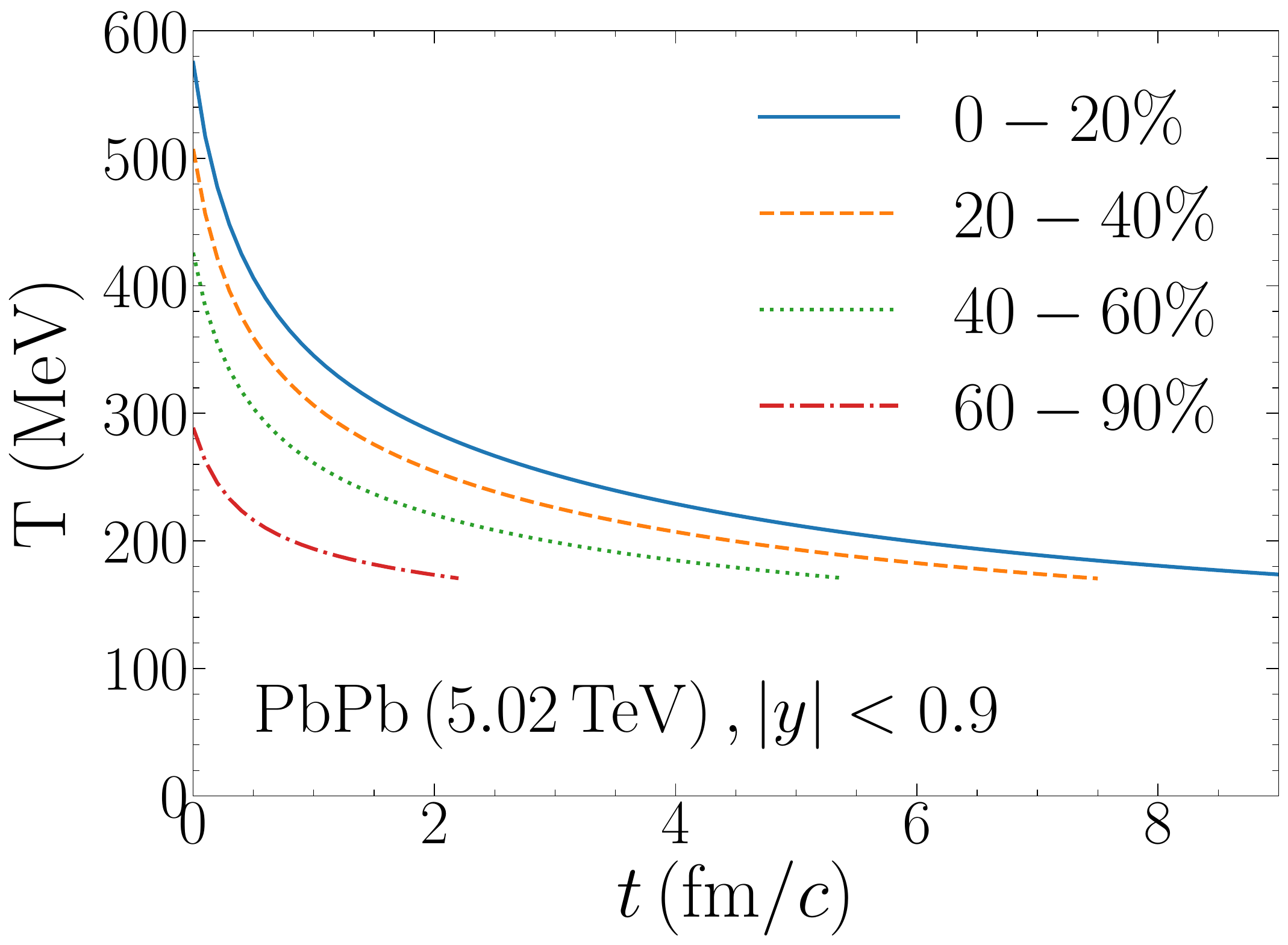}
\end{minipage}
\caption{Time evolution of temperature in the expanding fireball model in 5\,TeV Pb-Pb collisions at different centralities.} 
\label{fig_temp}
\end{figure}
This cross section includes an essentially 100\% feeddown contribution from strong and electromagnetic decays for all states below the hadronic $DB$ threshold of $m_D+m_B\simeq 7.15$\,GeV. This is different from charmonia and bottomonia where, \eg, the $P$ states can have significant branching ratios into hadronic final states through strong $Q\bar Q$
annihilation (which for $\bc$ states requires a weak interaction). To estimate the relative partition into $S$ and $P$ states, we take guidance from the corresponding 
production ratios in the charmonium and bottomonium cases. For the $\chi_c(1P)/J/\psi$ ratio, one has about
0.75~\cite{LHCb:2012af} and for the $\chi_b(1P)/\Upsilon(1S)$ about 1.08; thus, we estimate the $\bcp/\bcs\simeq 1$ in $pp$ collisions; \ie, half of the inclusive $\bcs$ arise from $1P$ feeddown.
%
This implies that the feeddown fraction of excited states to the inclusive $1S$ ground state production (excluding weak decays) is much larger than in the charmonium and bottomonium sector (cf.~also the pertinent discussion in Ref.~\cite{He:2022tod}). Moreover, since the $\bc$ ground-state meson observed thus far is most likely the pseudoscalar one (\ie, $\eta_c$-like), one can expect a 100\% feeddown from the slightly heavier vector state (through radiative decay) whose spin degeneracy is a factor of 3 larger.   
The shadowing of primordial $\bc$ production in Pb-Pb collisions is presumably between the shadowing of charmonia and bottomonia, thus, a 10-30\% shadowing is applied to the $\bc$ cross section~\cite{Armesto:2006ph}. 
Figure~\ref{fig_shadow} shows the shadowing as a function of $\npart$. Unless otherwise specified,
we display the shadowing uncertainty as an error band for our calculations throughout the remainder of this paper.

\section{Time Evolution and Inclusive $\bc$ Production}
\label{sec_prod}

The time evolution of the $\bc$ states in Pb-Pb collisions at 5.02\,TeV can be solved from Eq.~(\ref{rate-eq}) once the temperature profile is specified. Toward this end, we employ a cylindrical fireball with a longitudinal 
and transverse expansion of blast-wave type~\cite{Grandchamp:2003uw,Zhao:2011cv,Du:2015wha,Du:2017qkv}. 
The temperature evolution is obtained from an isentropically expanding volume where the total entropy is 
adjusted to the experimentally observed hadron production for a given centrality at a chemical-freezeout
temperature of $T_{\rm ch}$=160\,MeV, and expansion timescales from hydrodynamic models adjusted to experimental $p_T$ spectra of light hadrons at thermal freezeout
($T_{\rm fo}\simeq100$\,MeV for central collisions). Matching a lattice-QCD fitted equation of state (EoS) to a hadron resonance gas at 
$T_c=170\,$MeV~\cite{He:2011qa} results in the temperature evolution shown in Fig.~\ref{fig_temp}. Key parameters of the of the fireball evolution are summarized in Table~\ref{table:1}, specifically the initial longitudinal size $z_0$ (related to the formation time by the rapidity width, $\Delta y\simeq 1.8$, of the fireball, $z_0=\tau_0 \Delta y$), the longitudinal expansion velocity $v_z$ of the fireball cylinder's edges, the initial transverse radius, $R_0$, which depends on centrality, the transverse acceleration (implemented relativistically), and the total fireball entropy for the most central Pb-Pb collisions at 5\,TeV. 
\begin{table}[t]
\centering
\begin{tabular}{|c | c | c | c|}
 \hline
 $z_0\,\left( {\rm fm} \right)$ & 0.36 \\
 \hline
 $v_z\,\left( {\rm fm}/c \right)$ & 1.4\\
 \hline
 $a_z\,\left( {\rm fm}/c^2 \right)$ & 0\\
 \hline
 $R_0\,\left( {\rm fm}\right)$ & 3.2-6.8\\
 \hline
 $a_T\,\left( {\rm fm}/c^2 \right)$ & 0.1  \\
 \hline
 $S_{\rm tot}$(0-5\%) & 27000 \\
 \hline
\end{tabular}
\caption{Key parameters of the expanding blast-wave type cylinder used in this work.}
\label{table:1}
\end{table}

\begin{figure}[tbh]
\begin{minipage}[b]{1.0\linewidth}
\centering
\includegraphics[width=0.95\textwidth]{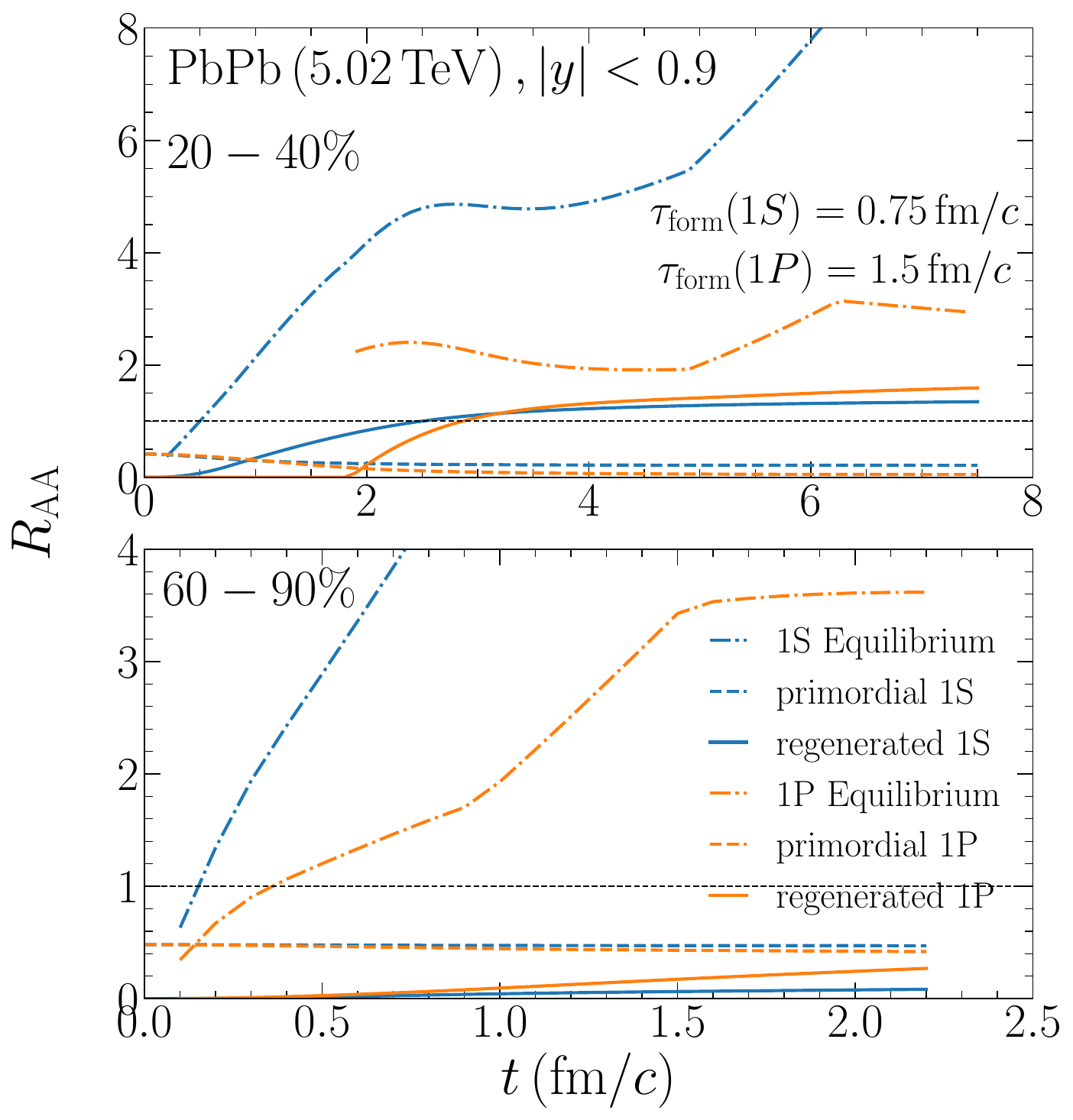}
\end{minipage}
\caption{Time evolution of the $\raa$'s of directly produced $\bc$ (1$S$: blue lines; 1$P$: orange lines, with primordial parts, regenerated parts, and equilibrium limit given by solid, dashed, and dot-dashed line styles, respectively) in 5\,TeV Pb-Pb collisions 
with an inclusive $pp$ production cross section of $\dd \sigma_{pp}^{\bc}/\dd y=91.5\,$nb.The upper (lower) panel shows the results for 20-40\% (60-90\%) centrality.
} 
\label{fig_time-evo}
\end{figure}
With the space-time evolution and initial conditions fixed, we solve the rate equation for the number of the individual $\bc$ states, $N_{Pb Pb}^{\bc}$. 
We recall the definition of the nuclear modification factor,
\begin{equation}
   \raa^{\bc}=\frac{N_{Pb Pb}^{\bc}(\npart)}{N_{\rm coll}(\npart) N_{pp}^{\bc}} \ ,
   \label{RAA-eq}
\end{equation}
which normalizes the yield in Pb-Pb to the one in proton-proton collisions, $N_{pp}^{\bc}$, scaled by the number of initial binary nucleon-nucleon collisions, $N_{\rm Coll}$.
We show the time evolution of the $\raa$'s for $S$- and $P$-wave $\bc$ in semi-central and peripheral Pb-Pb collisions at 5.02\,TeV at mid-rapidity in Fig.~\ref{fig_time-evo}. 
With 100\% feeddown from $\bcp$ decays, the inclusive $\bcs$ result amounts to the sum of the
$\raa$'s from the $S$- and $P$-wave states (using the inclusive yield in the denominator of all $\raa$'s). 
Both primordial parts are rather strongly suppressed in the early phases 
of the medium evolution in semi-central collisions, even though we include initial formation time effects, which suppress 
the reaction rates by a factor $\tau/\tau_{\rm form}$ 
for $\tau \leq \tau_{\rm form}$ to account for the expansion of a small-size $b\bar c$ pair into 
a fully formed bound-state (note that the scaling is linear in time, not quadratically as one
would expect from a classical cross section picture)~\cite{Farrar:1988me}, with 
$\tau_{\mathrm{form}}(1S)=0.75\,$fm/$c$, and $\tau_{\mathrm{form}}(1P)=1.5\,$fm/$c$. In peripheral collisions, the primordial component is much less suppressed due to the short fireball lifetime, but $\bcp$ is significantly more suppressed than $\bcs$ as a consequence of larger reaction rates for $\bcp$ at the lower temperatures.

Regeneration of the $\bc$ states commences when the medium has cooled down to their respective ``dissociation'' temperatures (with no regeneration operative before that). Following our previous applications to charmonia and bottomonia, we conservatively adopt dissociation temperatures at vanishing binding energy, \ie, $\Td(1S)$=420\,MeV and $\Td(1P)$=260\,MeV as indicated in Fig.~\ref{fig_BE}. One could also argue that the quantum mechanical uncertainty implies that bound states are only well-defined for binding energies of the order of the width or larger. However, even for small $E_B$, resonance-like correlations can persist which allow for the population of a pertinent quantum state. A more accurate description of this regime, as well as of
the formation time effect, requires a quantum-transport treatment. In semi-central collisions a large reaction rate (recall Fig.~\ref{fig_rate_Bc}) and a large degeneracy lead to a $\bcp$ contribution to the inclusive $\bcs$ yield that is comparable to (even slightly larger than) the direct $\bcs$ contribution. On the other hand, in peripheral collisions, both $\bc$ states start regenerating at 
almost the same time, but with substantially larger rates for the $\bcp$ resulting in a larger yield than for the $\bcs$.

\begin{figure}[!t]
\begin{minipage}[b]{1.0\linewidth}
\centering
\includegraphics[width=0.98\textwidth]{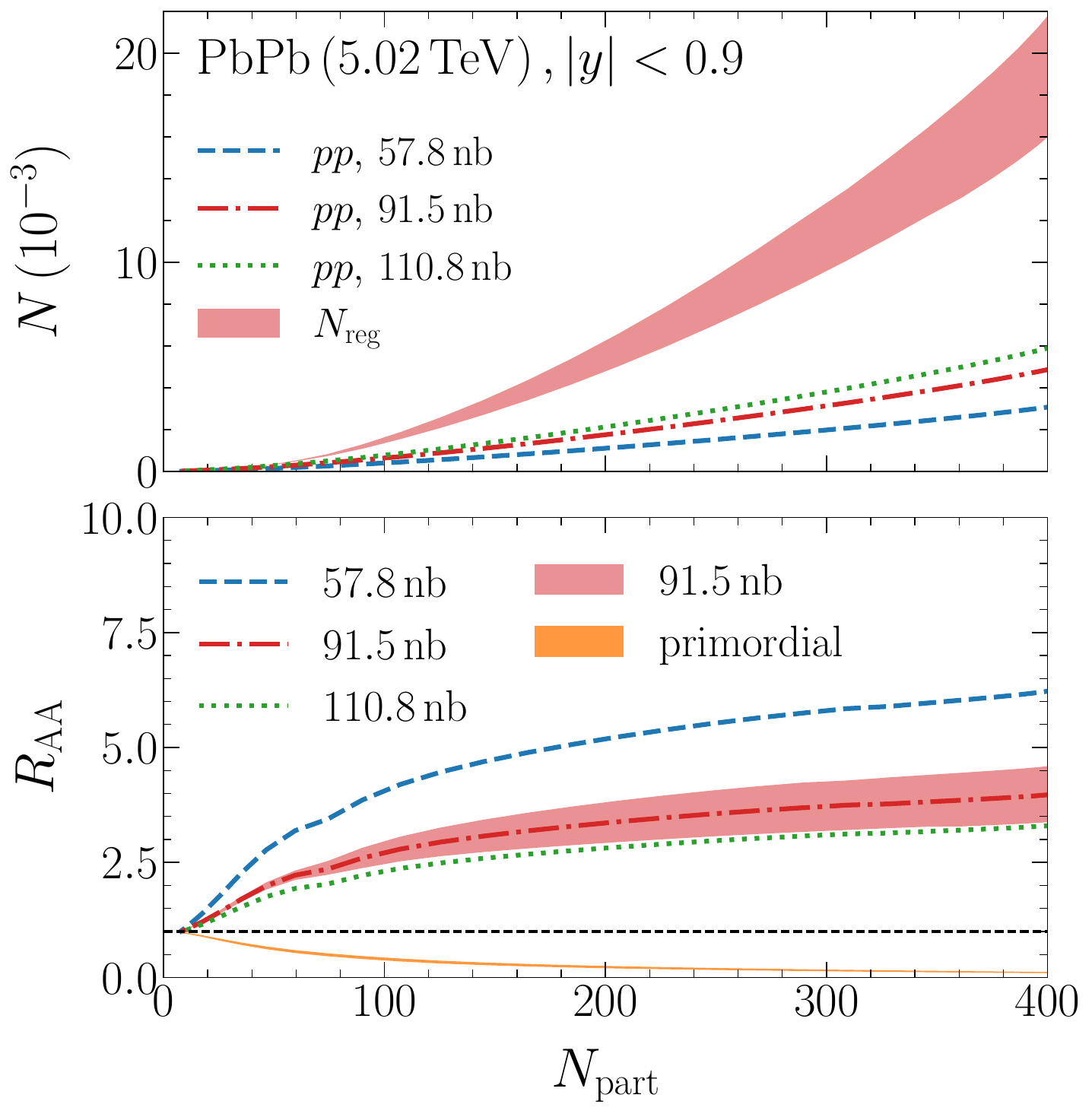}
\end{minipage}
\caption{Upper panel: Centrality dependence of the inclusive yield of regenerated $\bcs$ (red band) and the unsuppressed primordial production based on three different input cross sections in $pp$ collisions.   
Lower panel: $\raa$ of total (sum of suppressed primordial and regeneration) inclusive $\bcs$ production where the blue-dashed, red dot-dashed and green dotted lines correspond to $pp$ production cross-sections of 57.8, 91.5 and 110.8$\,$nb, respectively (figuring in both the denominator and the primordial component in the numerator), with 20\% (5\%) $c$ ($b$) quark shadowing and 20\% shadowing of the primordial $\bc$'s;
the solid orange line shows the primordial $\bcs$ $\raa$ component with a $pp$ production cross section of 91.5\,nb. In both panels the upper limit of the red bands correspond to 10\% (0) shadowing for $c$ ($b$) quarks and 10\% shadowing for primordial $\bc$'s, while the lower limits include 30\% (10\%) shadowing for $c$ ($b$) quarks and 30\% 
in the primordial part.}
 \label{fig_raa-npart}
\end{figure}
The centrality dependence of the regenerated $\bc$ yield in absolute terms is shown in the upper panel of Fig.~\ref{fig_raa-npart}, together with the unsuppressed primordial yields for the different $pp$ cross sections that we employ and that figure in the denominator of the $\raa$. Clearly, 
the inclusive-$\bc$ $\raa$ and its decomposition into regenerated and primordial parts is shown in Fig.~\ref{fig_raa-npart}. Even for rather peripheral collisions, the regeneration yield rapidly builds up to produce an $\raa$ that is well above one. 

\section{Transverse-momentum spectra}
\label{sec_pt}
%
\begin{figure}[t]
\begin{minipage}[b]{1.0\linewidth}
\centering
\includegraphics[width=0.96\textwidth]{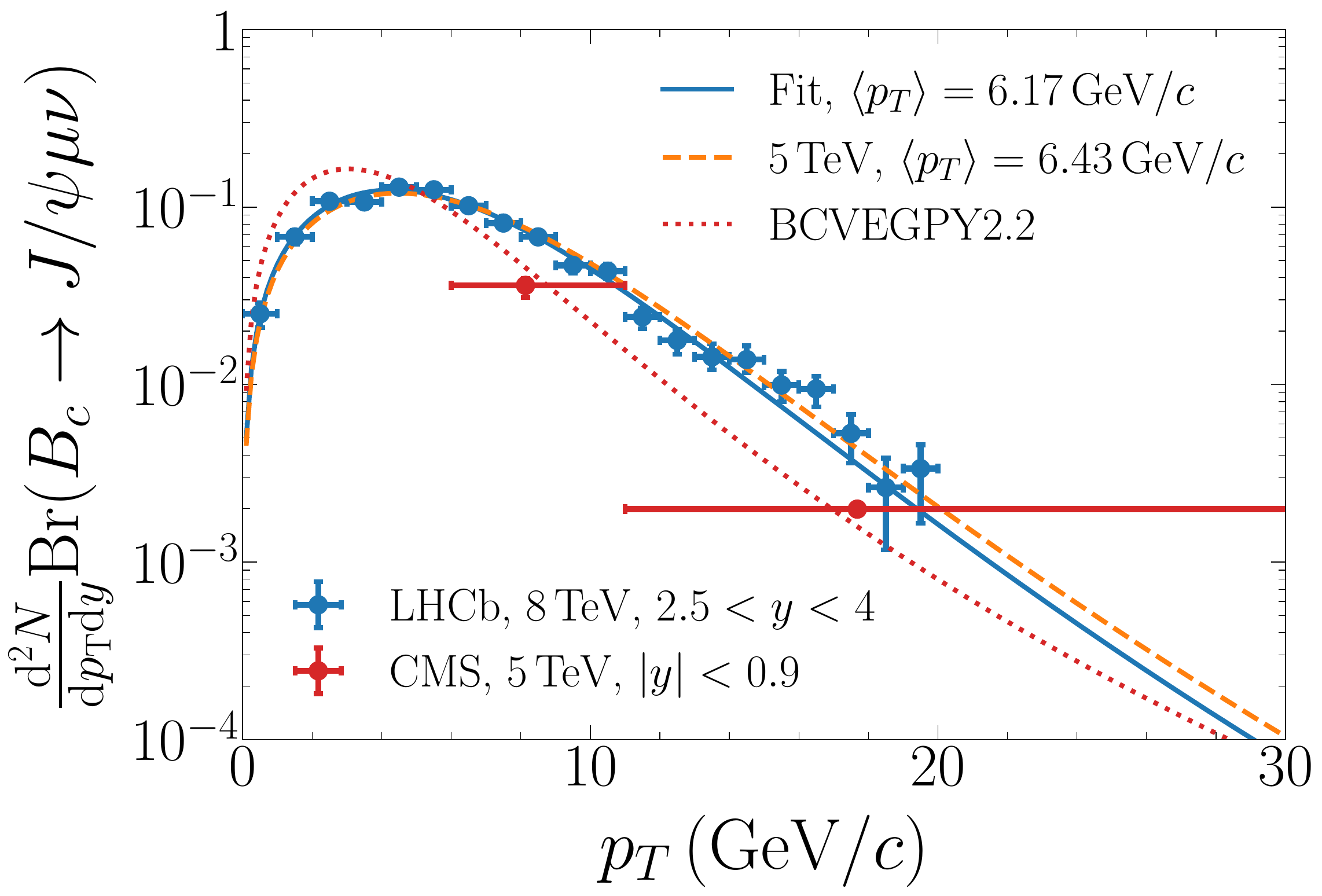}
\end{minipage}
\caption{Transverse-momentum spectra in $pp$ collisions for $\bc^+\to J/\psi\pi^+$ measured at 8\,TeV and forward-rapidity (blue dots)~\cite{LHCb:2014mvo} and $\bc^-\to J/\psi\mu^-\bar \nu$ at 5.02\,TeV and mid-rapidity (red dots)\cite{CMS:2022sxl}. 
The curves are fits to 8 TeV LHCb data (blue line), extrapolated to mid-rapidity at 5\,TeV (orange dashed line), and a fit to 5\,TeV CMS data based on the code package BCVEGPY2.2~\cite{Zhao:2022auq} (red dotted line). 
}
\label{fig_pt_pp}
\end{figure}
In this section we utilize our rate equation results to compute the $\pT$ spectra of $\bc$ mesons.
We first calculate the $\pT$ dependence of the primordial component that follows 
from a suppression calculation in a Boltzmann equation initialized by suitably constructed $\bc$ spectra in $pp$ collisions in Sec.~\ref{ssec_prim}. We then employ charm- and bottom-quark spectra that have been transported through the QGP using relativistic Langevin simulations~\cite{He:2014cla} with non-perturbative heavy-light $T$-matrix interactions and result in a fair phenomenology 
of HF hadron production at the LHC, to evaluate the $\pT$ spectra of regenerated $\bc$'s. 
Toward this end, we approximate recombination processes to occur at a fixed temperature 
representing an average of the continuous regeneration for each state using two different
recombination models, namely a widely used instantaneous coalescence in Sec.~\ref{ssec_icm} and resonance recombination in Sec.~\ref{ssec_rrm}.

%

\subsection{Initial $\pT$ Spectra and Primordial Component}
\label{ssec_prim}
\begin{figure}[t]
\begin{minipage}[b]{1.0\linewidth}
\centering
\includegraphics[width=0.9\textwidth]{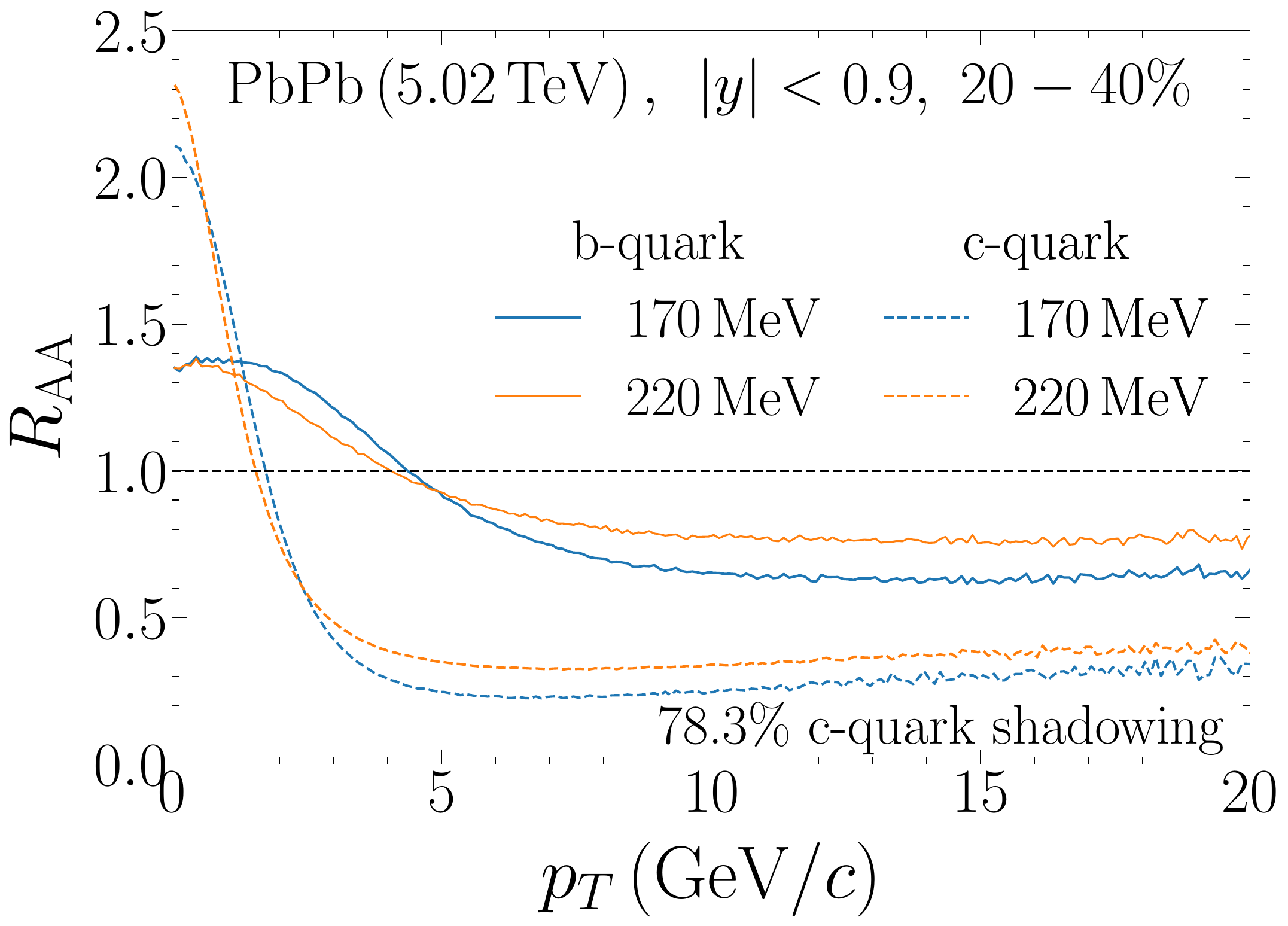}
\end{minipage}
\caption{Nuclear modification factor of $b$ quarks (solid lines) and $c$ quarks (dashed lines) at $T$=170\,MeV (blue) and $T$=220\,MeV (orange) as obtained from Langevin simulations through the QGP phase in 20-40\% central 5\,TeV Pb-Pb collisions.}
\label{fig_pt_quarks}
\end{figure}

The most accurate $p_T$ spectra of $\bc$ in $pp$ collisions to date are from the LHCb Collaboration at 8 TeV and forward rapidity ($2.0<y<4.5$)~\cite{LHCb:2014mvo}. We have fitted those using the ansatz
\begin{equation}
   \frac{\dd N_{pp}^{\bc}}{\dd \pT^2}=\frac{N}{\left(1+\left({\pT}/{A}\right)^2\right)^n} \ , 
   \label{p_T}
\end{equation}
obtaining $N=0.0078\pm0.0003$, $A=13.72\pm2.16$\,GeV and $n=5.62\pm1.40$, resulting in $\langle \pT \rangle=6.17\,$GeV for the central fit values; cf.~Fig.~\ref{fig_pt_pp}. We extrapolate these to 5.02\,TeV at mid-rapidity by correcting for the mean transverse momentum, $\langle p_{\rm T}\rangle$, 
using an average of experimental results for its energy dependence from charmonia at forward~\cite{ALICE:2017leg} and mid-rapidity~\cite{ALICE:2019pid} and for the energy and rapidity dependence from bottomonia~\cite{LHCb:2018psc,CMS:2022wfi}; we find that $\langle \pT \rangle$ increases by 10\% when 
going from forward to mid-rapidity, and decreases by 5.4\% when going from 8 to 5\,TeV collision energy, amounting to $\langle \pT \rangle=6.43\,$GeV for $\bc$ at 5.02\,TeV and $|y|<0.9$, which we accommodate by adjusting $A$ to 14.3\,GeV.  
Alternatively, we have fitted $\pT$ spectra of $\bc$ mesons from CMS in 5.02\,TeV $pp$ collisions~\cite{CMS:2022sxl} using BCVEGPY2.2 simulations~\cite{Zhao:2022auq}; the fit of the Eq.~(\ref{p_T}) to the latter yields $N=0.0147$, $A=7.88\,$GeV and $n=3.86$, corresponding to $\langle \pT \rangle=4.8\,$GeV. Both cases for the fits and data (all normalized to an integrated norm of one)  are shown in Fig.~\ref{fig_pt_pp}. Unless otherwise stated, we will use the LHCb-based fit in the denominator of the $\raa(\pT)$.

To compute the $\pT$ dependence from the transport model in Pb-Pb, we take advantage of a decomposition of the rate equation into primordial and regenerated components 
corresponding to its homogeneous and inhomogeneous solutions, respectively~\cite{Zhao:2007hh}. The $\pT$ dependence of the former is obtained by solving the Boltzmann equation for the $B_{c}$ distributions, $f_{\bc}$, as
\begin{equation}
    \begin{aligned}
   &f_{\bc}(\vec{x},\vec{p},\tau)\\
   &=f_{\bc}(\vec{x}-\vec{v}(\tau-\tau_0),\vec{p},\tau_0)
{\rm e}^{-\int\limits_{\tau_0}^{\tau}\Gamma_{\bc}(\vec{p},T(\tau'))\dd \tau'} 
\end{aligned}
\end{equation}
with an initial condition from the $pp$ spectra including shadowing~\cite{Du:2017qkv}. The homogeneous solution is subtracted from the inhomogeneous one to normalize the
$\pT$ spectra from regeneration which we calculate in the following using two different
recombination models.

\subsection{Instantaneous Coalescence}
\label{ssec_icm}
%
\begin{figure}[t]
\begin{minipage}[b]{01.00\linewidth}
\centering
\includegraphics[width=0.95\textwidth]{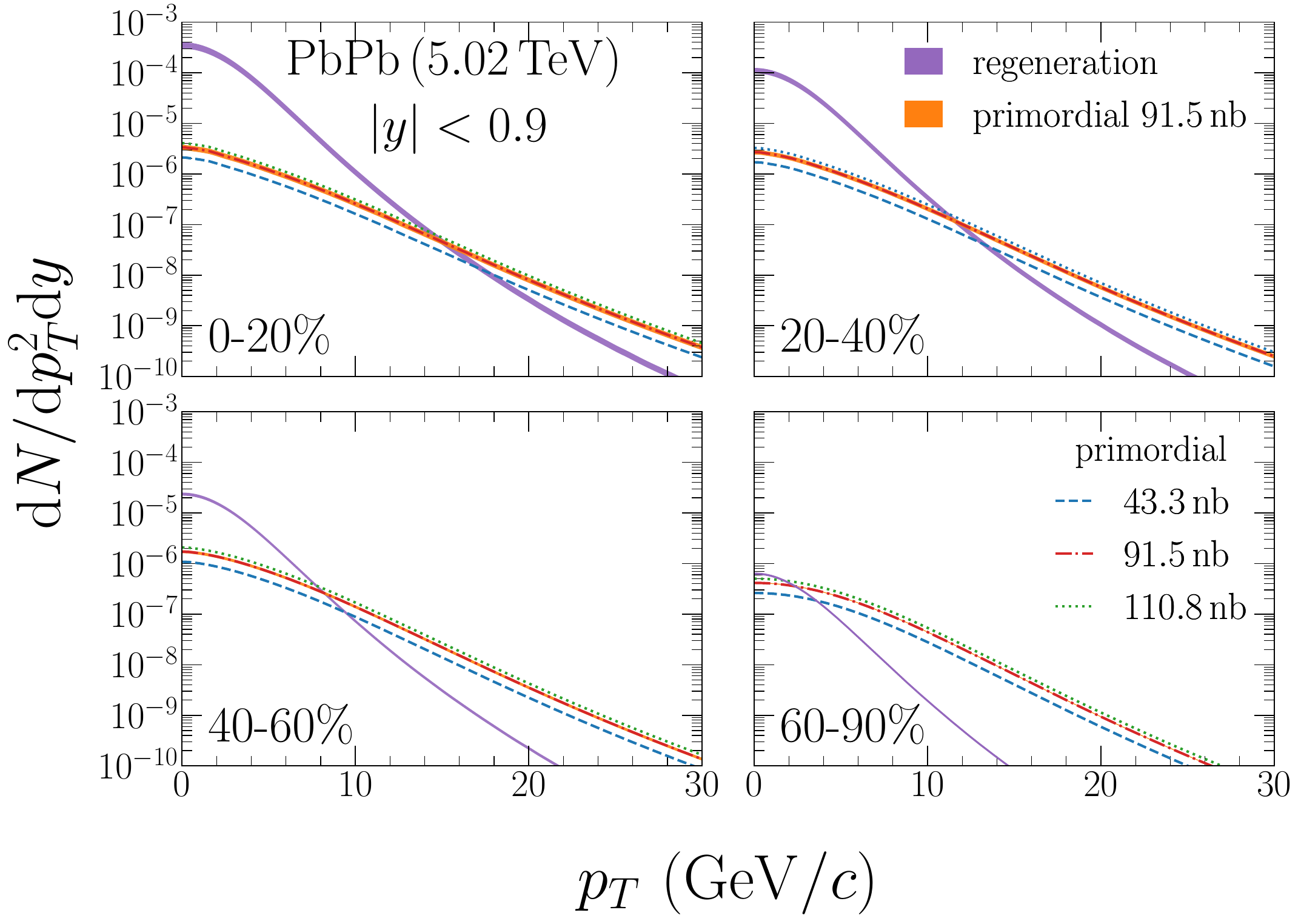}
\end{minipage}
\caption{The $\pT$ spectra of primordial (orange bands and lines for different $pp$ cross sections) and regenerated (purple) $\bcs$ production for different centralities in Pb-Pb(5.02\,TeV)
collisions using an ICM for recombination. 
}
\label{fig_Npt-icm}
\end{figure}

The ICM has been widely applied as a mechanism of hadronization in HICs~\cite{Fries:2008hs},
in particular for the explanation of the empirical ``constituent-quark number scaling" of the 
$v_2$ and large baryon-to-meson ratios for light-hadron production at ``intermediate"
$\pT\simeq$\,2-6\,GeV. It has also been applied to charm-quark hadronization 
~\cite{Greco:2003vf,Plumari:2017ntm}. Its main virtue is that it can account for 
off-equilibrium (non-thermalized) quark spectra. Here we apply it to $\bc$ mesons in a  standard form which assumes global quark distributions in coordinate space and is given by
\begin{equation}
\begin{split}
  &\frac{\dd^{3} N_{\bc}^{\mathrm{coal}}\left(\mathbf{p}\right)}{\dd^{3} \mathbf{p}}={C_{\mathrm{reg}}}g_{\bc}\int \dd^{3} \mathbf{p}_{c} \dd^{3} \mathbf{p}_{\barb} \frac{\dd^{3} N_{c}}{\dd^{3} \mathbf{p}_{c}} \frac{\dd^{3} N_{\barb}}{\dd^{3} \mathbf{p}_{\barb}}
\\
&\!\times\delta^{(3)}\left(\mathbf{p}-\mathbf{p}_{c}-\mathbf{p}_{\barb}\right)w\left( \mathbf{k} \right)\ .
\label{coal-eq}
\end{split}
\end{equation}
{The initial-state averaged and final-state summed degeneracy factors are $g_{\bc}=1/9$ and $1/3$ for $\bcs$ states (with total spin degeneracy 4) and $\bcp$ states (with total spin degeneracy 12), respectively, accounting for the probability of
forming a colorless meson of given spin from the underlying quark color and spin.} The coalescence probability 
of the $c$ and $\bar b$ quarks is encoded in the Wigner 
distribution~\cite{Sun:2017ooe}, 
\begin{equation}
\begin{aligned}
w(\mathbf{k}) &=\frac{\left(4 \pi \sigma^{2}\right)^{\frac{3}{2}}}{V_{\rm FB}} \frac{\left(2 \sigma^{2} \mathbf{k}^{2}\right)^{l}}{(2 l+1) ! !} e^{-\sigma^{2} \mathbf{k}^{2}} \ ,
\end{aligned}
\end{equation}
for a $\bc$ with angular momentum $l$; $k$ denotes the relative momentum of the two quarks,
the $\sigma$ are estimated from the mean-square radii of $\bc(1S)$ and $\bc(1P)$,
{and $V_{\rm FB}$ is the volume of the fireball. The factor in the Wigner distribution is introduced so that it satisfies $\int \dd^3\mathbf{x}\dd^3\mathbf{k} w\left(k\right)= (2\pi)^3$~\cite{Greco:2003mm}}.
For different quark masses, $m_c$ and $m_{\bar b}$, one has~\cite{Sun:2017ooe}
\begin{equation}
\begin{aligned}
\mathbf{k}&=\sqrt{2}\frac{m_{\barb} \mathbf{p}_{c}-m_c \mathbf{p}_{\barb}}{m_c+m_{\barb}}\\
\sigma^2\left( 1S \right)&=\frac{2}{3}\frac{\left( m_c+m_{\barb} \right)^2}{m_c^2+m_{\barb}^2}\langle r^2_{1S}\rangle\\
\sigma^2\left( 1P \right)&=\frac{2}{5}\frac{\left( m_c+m_{\barb} \right)^2}{m_c^2+m_{\barb}^2}\langle r^2_{1P}\rangle \ .
\end{aligned}
\end{equation}
Note that there is a Jacobian factor
$w\left( \mathbf{p}_c,\mathbf{p}_{\bar{b}}\right)=2^{\frac{3}{2}}w\left( \mathbf{k}\right)$.

{We employ Eq.~(\ref{coal-eq}) at midrapidity to obtain the $\pT$ spectra of $\bc$ at $p_{z}=0$:
$\frac{\dd^{3} N_{\bc}^{\mathrm{coal}}\left(\mathbf{p}_T\right)}{\dd y \dd^{2} \mathbf{p}_T}= E_{\bc} \frac{\dd^{3} N_{\bc}^{\mathrm{coal}}\left(\mathbf{p}\right)}{\dd^{3} \mathbf{p}} \vert_{p_z = 0} $  with $\dd p_{z}=E_{\bc} \dd y$.} 
The normalization constant, $C_{\rm reg}$, introduced above, {is about 1.2  for the $1S$ state and $\approx$ 0.5 for the $1P$  state in central collisions, and depends on the values chosen for the 
radii of $\bc$, $r_{1S[1P]} =0.35[0.75]$\,fm, assumed} to lie in between the radii
of charmonia and bottomonia~\cite{Du:2017qkv}.
For the HQ spectra we employ the results of relativistic Langevin simulations 
(shown in Fig.~\ref{fig_pt_quarks})~\cite{He:2012xz} at $T$=220\,MeV 
as an average temperatures, which we use for simplicity for both $\bcs$ and $\bcp$ regeneration production (we have checked that using the HQ $\pT$ spectra for the $\bcp$ at, \eg, $T$=170\,MeV, leads to a maximal modification of less than 20\% in the regenerated $\bc$ $\raa$ around 15\,GeV).
\begin{figure}[t]
\begin{minipage}[b]{01.00\linewidth}
\centering
\includegraphics[width=0.9\textwidth]{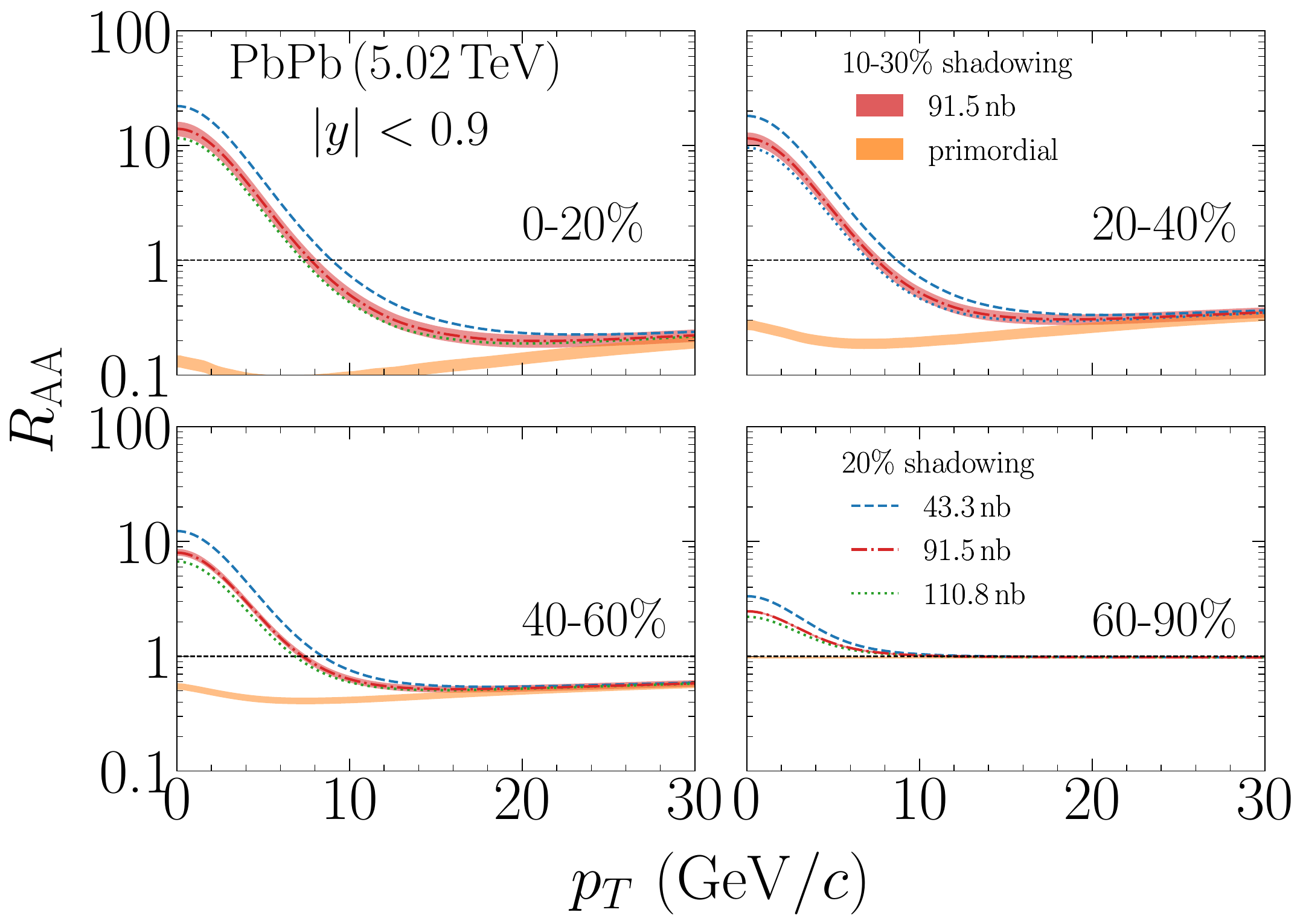}
\end{minipage}
\caption{The $\bcs$ $\raa$ for primordial (orange bands) and regenerated (red bands for $d\sigma/dy$=91.5\,nb with shadowing uncertainty, and lines for other $pp$ cross sections with fixed shadowing) production as a function of $\pT$ for different centralities in Pb-Pb(5.02\,TeV)
collisions using an ICM for recombination. 
}
\label{fig_raa-icm}
\end{figure}

\begin{figure}[!t]
\begin{minipage}[b]{01.00\linewidth}
\centering
\includegraphics[width=0.98\textwidth]{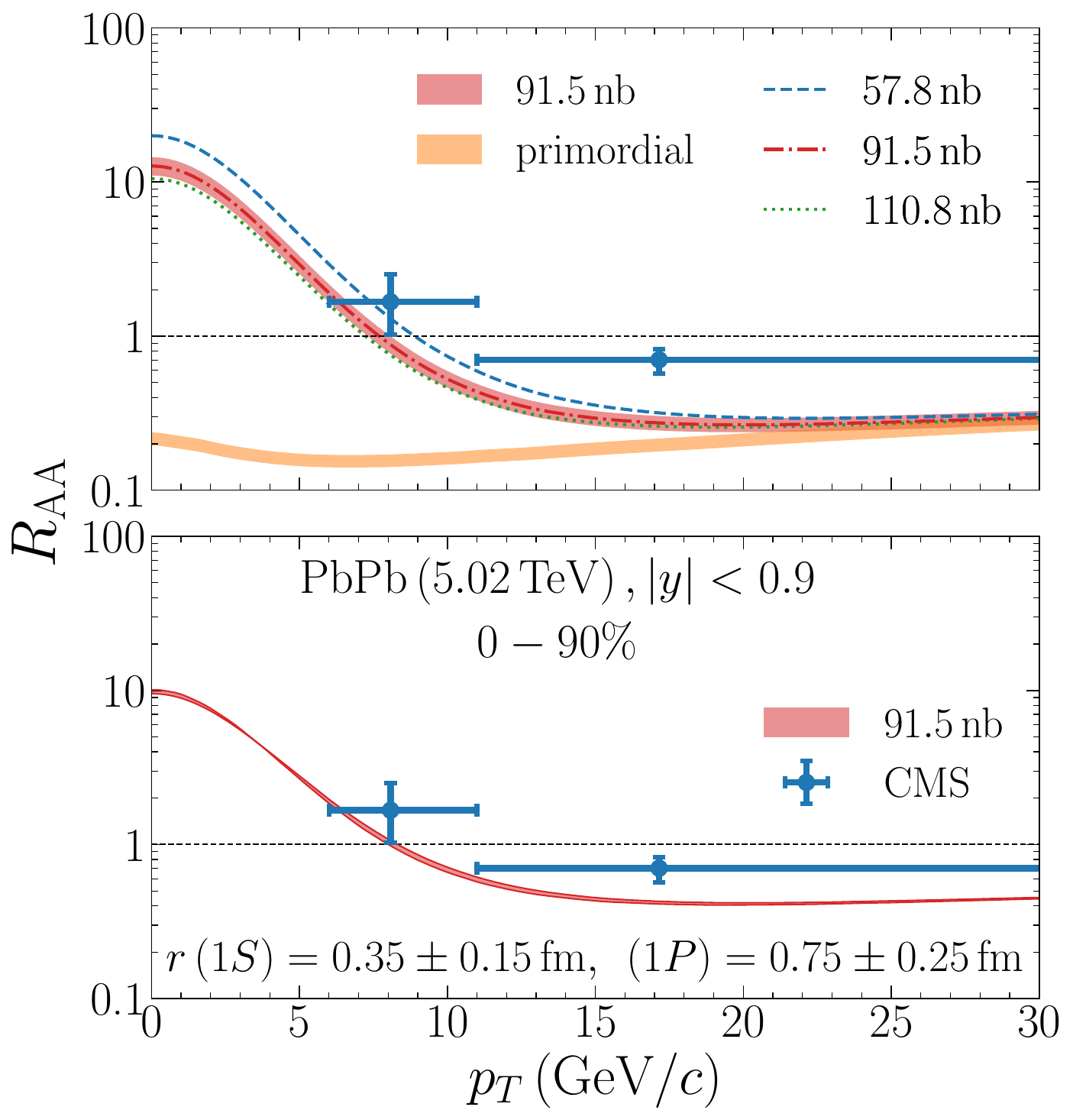}
\end{minipage}
\caption{The $\raa$ of inclusive $\bc^\pm$ vs. $p_{\rm T}$ in 0-90\% 5.02\,TeV Pb-Pb
collisions using an ICM for recombination. The bands and lines in the upper panel have the same meaning as in Fig.~\ref{fig_raa-npart}. The lower panel illustrates 
uncertainties due to a variation of the $\bc$ radii in the ICM. Our calculations
are compared to CMS data~\cite{CMS:2022sxl}.
}
\label{fig_pt_0-90}
\end{figure}

The ICM results are combined with the suppressed primordial component to calculate the (absolutely normalized) $\pT$ spectra and the nuclear modification factor, $\raa(\pT)$, for inclusive $\bcs$  production in various centrality bins in 5.02\,TeV Pb-Pb collisions, see Figs.~\ref{fig_Npt-icm} and \ref{fig_raa-icm}, respectively. In central collisions, the regeneration contribution dominates over the primordial one out to $\pT\simeq20$\,GeV, and is still quite noticeable in peripheral collisions at low $\pT$.
In the $\raa$'s, the uncertainty of the $pp$ cross section figuring in the denominator is larger than that from shadowing corrections.

In Fig.~\ref{fig_pt_0-90} we compare our inclusive $\bcs$ $\raa(\pT)$ for 0-90\% central collisions (obtained from the centrality bins in Fig.~\ref{fig_raa-icm}) to CMS data~\cite{CMS:2022sxl}.
Again, the total $\raa$ is dominated by the regeneration component out to momenta of around $\pT\simeq15$\,GeV, reaching large values of 10 or more at low $\pT$. 
The upper and lower panels 
illustrate, respectively, uncertainties due to the $pp$ input cross section 
and the coalescence radii of $\bc$, $r_{1S[1P]} =0.35[0.75]\pm 0.15[0.25]$\,fm.
While the former are large, the latter are comparatively small. 


\subsection{Resonance Recombination}
\label{ssec_rrm}
\begin{figure}[t]
\begin{minipage}[b]{01.00\linewidth}
\centering
\includegraphics[width=0.98\textwidth]{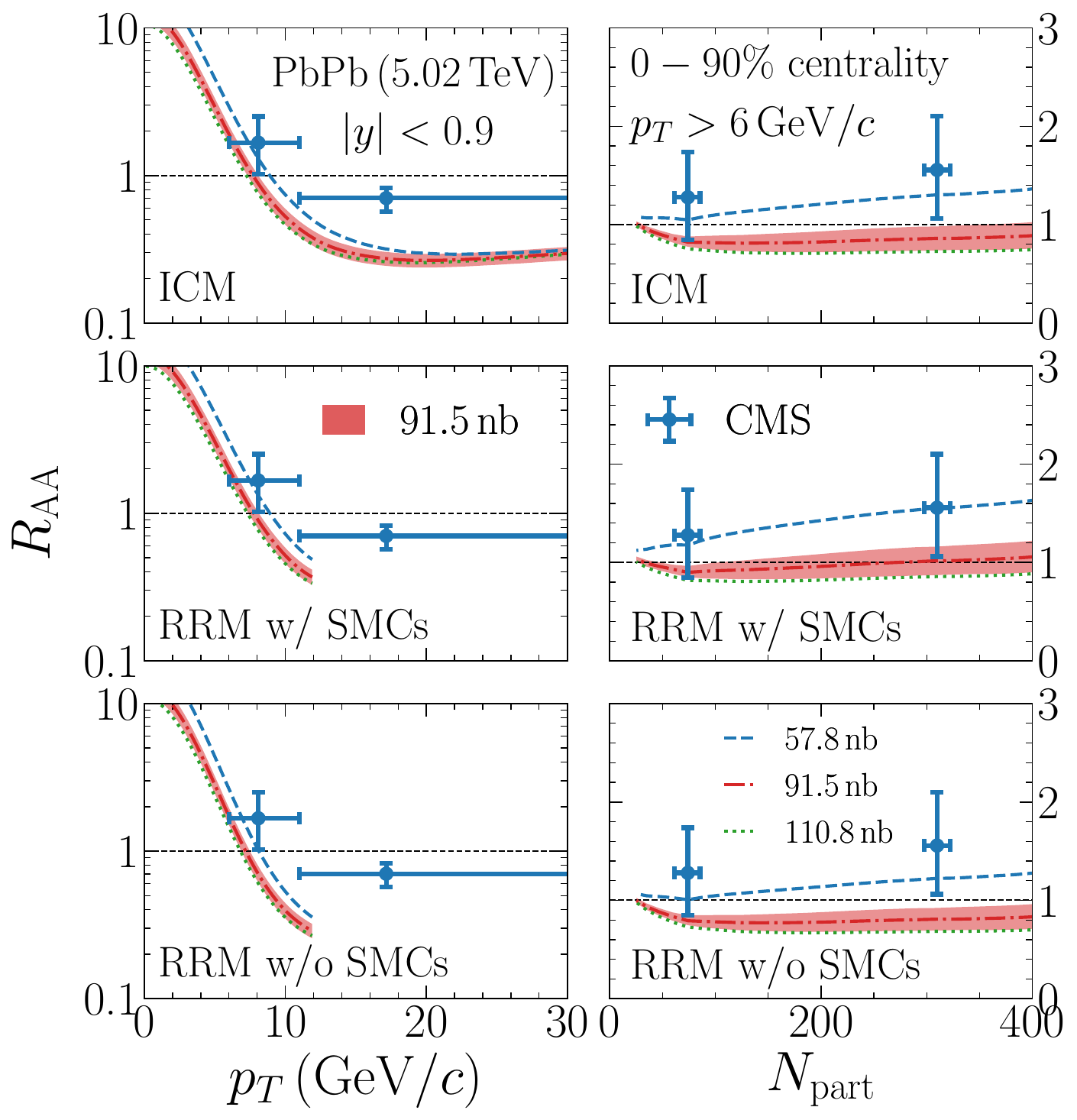}
\end{minipage}
\caption{
   The $\raa$ of inclusive $\bc^\pm$ production vs.~$\pT$ (left column) and $\npart$ with $\pT>6\,$GeV (right column) in 5.02\,TeV Pb-Pb collisions with 
   regeneration from ICM (upper panels), RRM with (middle panels) and without SMCs (lower panels), compared to CMS data~\cite{CMS:2022sxl}.
The $pp$ reference spectra are from LHCb~\cite{LHCb:2014mvo}. The bands and lines have the same meaning as in Fig.~\ref{fig_raa-npart}. 
}
\label{fig_lhcb}
\end{figure}

To assess the model dependence of the $\pT$ spectra of the regeneration component, we have conducted calculations 
using the resonance recombination model (RRM)~\cite{Ravagli:2007xx} which conserves 
four-momentum and recovers the equilibrium limit for equilibrated HQ input distributions
(also in the presence of radial and anisotropic medium flow)~\cite{He:2011zx}. More recently, the RRM has been extended to incorporate space momentum correlations (SMCs) between the coalescing quarks~\cite{He:2019vgs,He:2021zej}, which, \eg, enhance the recombination of fast-moving heavy quarks with high-flow thermal quarks in the outer regions of the fireball. Here, they
pertain to the diffusing $b$ and $\bar{c}$ quarks. 
The current implementations of the RRM, applied on a hydrodynamic hypersurface, also require an overall normalization constant, { typically of the order of 5} (which roughly corresponds to the number of re-generations when computed over a finite time interval).

The resulting $\raa$'s for the
regeneration component with and without SMCs, and for different $\bc$ cross 
sections in $pp$ collisions, are shown in Fig.~\ref{fig_lhcb}. For the $\pT$ dependence (left column), the spectra with SMCs are harder than the spectra without SMCs (although not by much), while the former are surprisingly close to the
ICM results in the upper left panel in Fig.~\ref{fig_lhcb}. In the right column, we present the centrality dependence by integrating the $\pT$ spectra over $\pT>6\,$GeV and compare to CMS data~\cite{CMS:2022sxl}. Smaller $pp$
input cross sections for $\bc$ production tend to give a better description of the data, in particular toward higher $\pT$.

\begin{figure}[t]
\begin{minipage}[b]{01.00\linewidth}
\centering
\includegraphics[width=0.95\textwidth]{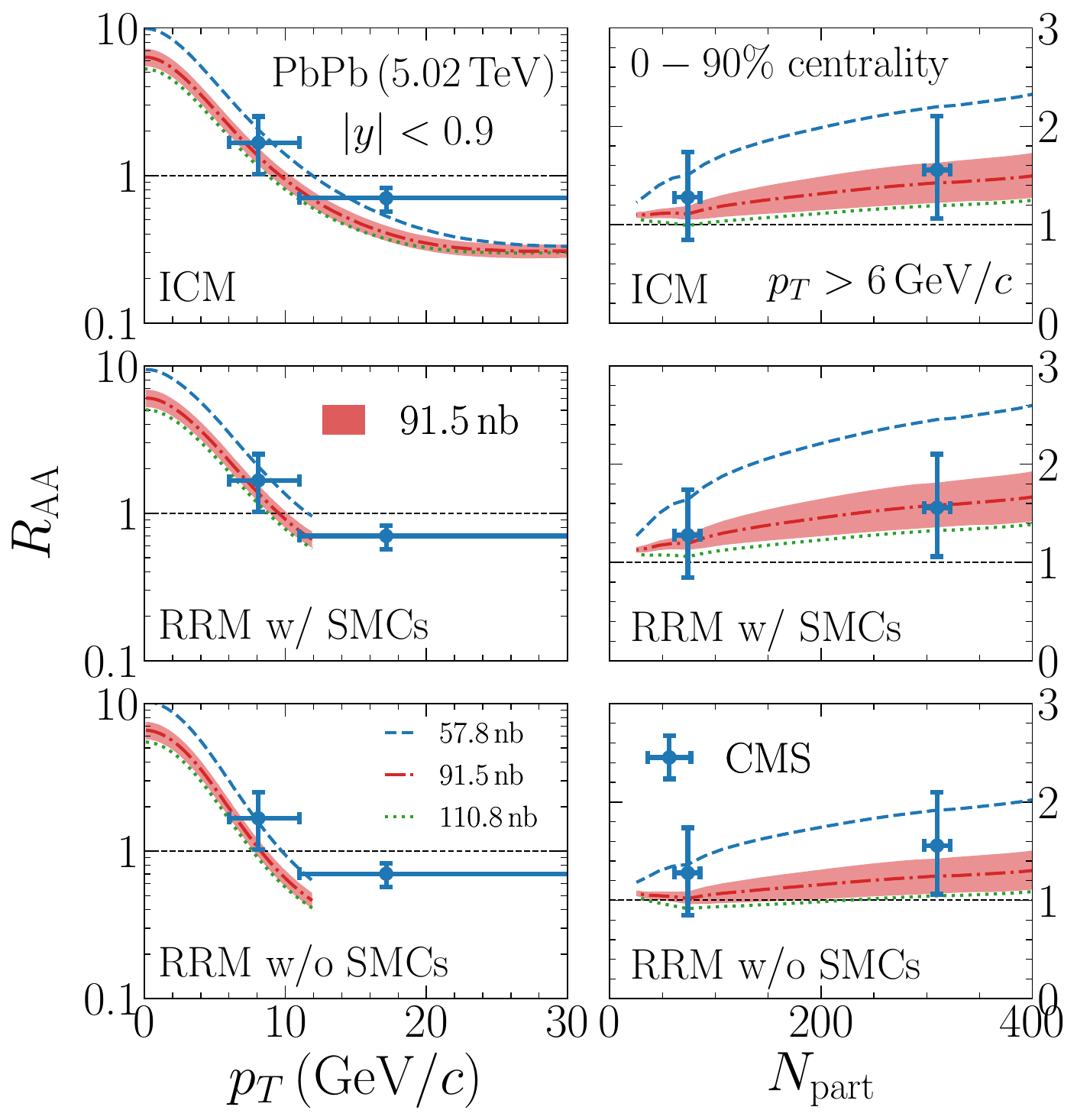}
\end{minipage}
\caption{
   Same as in Fig.~\ref{fig_lhcb}, but with CMS $pp$ $\pT$ spectra~\cite{CMS:2022sxl} in the denominator of the $\raa$.  
}
\label{fig_cms}
\end{figure}
In our comparison to CMS data shown above we have used $\bc$ $\pT$ spectra in $pp$ 
obtained from our extrapolation of a fit to LHCb data. However, the CMS Pb-Pb data for the $\raa(\pT)$ are based on a fit to the CMS $pp$ spectra. {Therefore, we display in Fig.~\ref{fig_cms} the results when using our fit to the CMS $pp$ data as shown in Fig.~\ref{fig_pt_pp}.} While this does not affect the primordial contribution to the $\raa$, the softer $\pT$ dependence of this fit implies a significant increase of the coalescence portion at higher $\pT${, and therefore the CMS Pb-Pb data for $\raa(\pT)$ are better described with lower values for the $pp$ input cross section,} at least for the ICM and RRM with SMCs. This reiterates the importance of an accurate experimental measurement of this quantity.




\section{Conclusions}
\label{sec_concl}
We have investigated the production of $\bc$ mesons in heavy-ion collisions using a thermal-rate equation approach. We first calculated $\bc$ spectral functions in the QGP from a thermodynamic $T$-matrix and used them to extract the 
pertinent binding energies. The latter were implemented in the evaluation of dissociation rates from inelastic scattering of thermal partons off the $b$ and $c$ quarks in the $\bc$ mesons. We also constructed the $\bc$ equilibrium limits through a combination of $b$- and $c$-quark fugacities and included effects of incomplete HQ thermalization. We solved the rate equations for $1S$ and $1P$ states in Pb-Pb(5.02\,TeV) collisions and computed the centrality dependence of inclusive $\bc^\pm$ yields. With $\approx$ 100\% feeddown from excited states below the open HF threshold, the latter make up $\approx$ 50\% of the inclusive yield of the pseudoscalar $\bcs$ meson. Large regeneration contributions cause a markedly rising $\raa$ with centrality, reaching values of up to $\approx$ 4-6 in central collisions. We then calculated $\pT$ spectra of the $\bc$ using two 
different recombination models (ICM and RRM with and without SMCs). The spectra of $c$ and $b$ quarks used in this calculation were generated from relativistic Langevin transport simulation, and the pertinent $\pT$ spectra from regeneration were normalized to the yields from the rate equation. The inclusive $\pT$-dependent $\raa$ for the $\bc$ is dominated by regeneration contributions for $\pT\lsim$~10-15\,GeV in semicentral and central collisions, reaching values up to around 10 at low $\pT$. The primordial yield dominates in peripheral (central) collisions for $\pT\gsim 10(20)$\,GeV. The results for the $\raa$ are rather sensitive to the $\bc$ production cross section in 
$pp$ collisions. A more precise measurement of this quantity, and of the production systematics of $\bc$'s in heavy-ion collisions (with a potentially spectacular enhancement at low $\pT$), will provide unprecedented insights into their in-medium properties and a valuable complement to, and interface of, the charmonium and bottomonium sectors. 

\acknowledgments
This work was supported by the U.S. National Science Foundation under Grant No. PHY-1913286 and No. PHY-2209335, by the TAMU Cyclotron Institute's Research Development (CIRD) program, and by the U.S. Department of Energy, Office of Science, Office of Nuclear Physics through the Topical Collaboration in Nuclear Theory on ``Heavy-Flavor Theory (HEFTY) for QCD Matter'' under Award No.~DE-SC0023547.
One of us (M.H.) was supported by the NSFC under Grant No.~12075122.

\bibliography{refcnew}

\end{document}